\documentstyle[aps,multicol,epsfig,epsf,psfrag]{revtex}

\begin{document}

\title{ Simulation of the Sedimentation of a Falling Oblate}
\author{F. Fonseca and H. J. Herrmann}
\address{ICA1, University of Stuttgart, Pfaffenwaldring 27, 70569
Stuttgart, Germany. }

\date{\today} 
\maketitle 
\widetext
\begin{abstract}
We present a numerical investigation of the dynamics of one falling oblate ellipsoid 
particle in a viscous fluid, in three dimensions, using a constrained-force technique 
\cite{Kai}, \cite{Kaih} and \cite{Esa}. We study the dynamical behavior of the oblate 
for a typical downward motion and obtain the trajectory, velocity, and orientation of 
the particle. We analyze the dynamics of the oblate generated when the height of the 
container, the aspect-ratio, and the dynamical viscosity are changed. Three types of 
falling motions are established: steady-falling, periodic oscillations and chaotic 
oscillations. In the periodic regime we find a behavior similar to the case of falling 
flat strips reported in ref. \cite{Belmonte}. In the chaotic regime the trajectory of 
the oblate is characterized by a high sensitivity to tiny variations in the initial 
orientation. The Lyapunov exponent is $\lambda = 0.052 \pm 0.005$. A phase space 
comparing to the results of ref \cite{Nori}, is shown.
               
\end{abstract}
             

\begin{multicols}{2}
\narrowtext

\section{Introduction}
The way in which objects    fall to the ground has been studied since antiquity.
Objects were thought to return to ``their natural'' places by the ancient Greeks.
Newton showed that the bodies fall on earth driven by a constant acceleration.
But despite gravity's undeniable attraction, not all falling objects travel downwards
on straight trajectories. The tree leaves flutter to the ground in the autumn, 
exhibiting a complex motion and refusing to follow the shortest path.
       
A deep understanding of the motion of falling objects in a fluid is of great 
technical importance, and has been investigated in a variety of contexts, including 
meteorology \cite{Meteorol}, aircraft stability \cite{Mises}, power generation 
\cite{Power}, chemical engineering \cite{Chemique}, etc. Also Newton observed the 
complex motion of objects falling in both air and water \cite{Viets}. This phenomenon 
was also studied by Maxwell, who discussed the motion of a falling paper 
strip \cite{Maxwell}. 

In the nineties, Aref and Jones \cite{Aref} found through numerical solutions of 
Kirchhoff's equations that the trajectory for an object moving through an 
incompressible inviscid and irrotational fluid, is chaotic. Tanabe and Kaneko 
\cite{Tanabe}, using a phenomenological model for the falling of a one-dimensional 
(1D) piece of paper, including lift and dynamical viscosity, but neglecting the inertia 
of the fluid, describe five falling regimes. Two of them are chaotic. In 1997 Stuart 
B. Field et al. \cite{Nori}, investigated experimentally the behavior of falling  
disks in a fluid, and identified different dynamical regimes as  
function of the moment of inertia and the Reynold's number. They obtained experimental 
evidence for chaotic intermittency \cite{intermittency}. 
In 1998 Andrew Belmonte, et al. \cite{Belmonte}, in an experiment with thin flat strips 
falling through a fluid, observed only two motions: side to side oscillation (flutter) 
and end-over-end rotation (tumbling). They proposed a phenomenological model including 
inertial drag and lift which reproduces this motion, and yields the Froude similarity,
which describes the transition from flutter to tumble regime.
Mahadevan et al \cite{Maha}, in 1999 made an experiment of dropping horizontal cards
of thickness $d$ and width $w$, showing that the tumbling frecuency $\Omega$ scales 
as $\Omega \approx d^{1/2}w^{-1}$, consistent with a dimensional argument that 
balances the drag against gravity.

Given the difficulties to study this problem theoretically and experimentally, 
we took a computational approach simulating the falling of one oblate ellipsoid in
a viscous fluid in a three dimensional container. An oblate is an ellipsoid for 
which the two largest principal radia are equal. We organize the paper in the 
following manner. In section 2 we give an review over the model that we use. In Sec. 
3A we describe the main features of the falling oblate. In Sec. 
3B-D we present the results for the change in the initial height, the oblate's 
aspect-ratio and the dynamical viscosity, for the steady-falling regime. In sec. 
3E we show the periodic behavior and compare to the results of reference 
\cite{Belmonte}. In Sec. 3F the sensitivity to tiny variations in the oblate's 
initial orientation is presented in the chaotic regime. In sec. 3G the parameter 
phase space is sketched and compared to the results of reference \cite{Nori}. 
Section 4 summarizes present results and discusses possible further applications.    
 

\section{ Model}
The general idea, proposed by Fogelson and Peskin \cite{Fogelson}, is to work
with a simple grid for the resolution of the fluid motion at all times and represent 
the particles not as boundary conditions to the fluid, but by a volume force term 
or Lagrange multipliers in the Navier Stokes equations.
  
This technique was developed in the work of Kuusela, et al \cite{Esa}, Wachmann, et 
al \cite{Wachmann} and Hoefler et al. \cite{Kai}, \cite{Kaih}, and employs a numerical 
solver for the dynamical simulation of three-dimensional rigid particles in a Newtonian 
fluid, bounded by a rectangular container.

The motion of the fluid is described by the dimensionless Navier-Stokes equations:
\begin{equation}
\frac{\partial \vec{v}}{ \partial t} 
+(\vec{v} \cdot \nabla)\vec{v}= -\nabla p +\frac{1}{Re} \nabla^2\vec{v}+\vec{f}   
\end{equation}
Here $p$ and $\vec{v}$ are the pressure and the velocity of the fluid, respectively, 
and $\vec{f}$ is an external force.

The Reynolds number $Re$ is defined as 
\begin{equation}
Re:=\frac{UD\rho_f}{\nu}
\end{equation}

\noindent where $U$ is the mean vertical oblate velocity, $D$ a characteristic length that 
in our case is the largest oblate's diameter, $\rho_f$ the density and $\nu$ the 
dynamical viscosity of the fluid.

For an incompressible fluid, the continuity equation:
\begin{equation}
\frac{\partial \rho_f}{ \partial t}+\nabla\cdot(\rho_f\vec{v})=0
\end{equation}

reduces to
\begin{equation}
\nabla \cdot \vec{v}= 0   
\end{equation}
Equation (1) is discretized on a regular, marker-and-cell mesh to second order
precision in space. For the time stepping, we employ an operator-splitting-technique
which is explicit and accurate to first order. The suspending fluid is subjected to  
no-slip boundary conditions at the surface of the suspended particles. More details 
of the solution procedure are presented in  \cite{Esa}, \cite{Kai}, \cite{Kaih}. 

The geometry of the oblate ellipsoid is characterized by $\Delta r$ its aspect-ratio 
defined as the ratio of the smallest radius over the largest one.

An oblate is represented by a rigid template connected to fluid tracer particles, 
which are moving on the trajectories of the adjacent fluid. The connection is made by 
using the body force term, in the Navier-Stokes equations, as constraints on the fluid 
such to describe the oblate.

The force density $\vec{f}^c$, is chosen elastic with a spring constant that, 
and guarantees that the elongation remains small against the grid spacing at all times 
\cite{Wachmann}, and it is zero in the exterior of the region outside the oblate. We 
can define this force density $\vec{f}^c$ as:

\begin{equation}
\vec{f}^c=f^c(\vec{x_{ij}}+\vec{\epsilon}(\vec{x_{ij}})) = -k\vec{\epsilon}(\vec{x_{ij}})
\end{equation}

\noindent where $\vec{x_{ij}}$ is the displacement field of the separation between the markers 
$i$ and their corresponding reference point $j$. The stiffness constant $k$, must be chosen 
large enough so that $\mid\vec{\epsilon}(\vec{x_{ij}})\mid  \ll h $, ($h$ size grid), 
holds for all iterations.
 
In general the displacement field $\vec{\epsilon}(\vec{x_{ij}})$ is defined as:

\begin{equation}
\vec{\epsilon_i}(\vec{x_{ij}}^m)=\vec{x_{ij}}^m-\vec{x_{ij}}^r
\end{equation}

The vector $\vec{x_{ij}}^m$ is the position of a fluid tracer, whose motion is 
determined by the fluid local velocity, i.e.,

\begin{equation}
\dot{\vec{x_{ij}}^m}=\vec{u}(\vec{x}_{ij}^m)
\end{equation}

The $\vec{x_{ij}}^r$ are the reference points associated to a template having 
the shape of the physical particle:

\begin{equation}
\vec{x_{ij}}^r=\vec{x}_i+O_i(t).\vec{r}_{ij}
\end{equation}

Here $\vec{x}_i$ is the center of mass of the template, $O_i(t)$ is the rotation  
matrix that describes the present orientation of the oblate and $\vec{r_{ij}}$ denote 
the initial position of the reference points with respect to the center of mass. For the 
quaternion formulation of the rotation, we use the techniques described in ref. 
\cite{Quater}.
 
A velocity-Verlet integrator \cite{Integra} serves to integrate the equations of 
motion for the translation and a Gear-predictor integrator \cite{Allen} for the 
rotation on the template:

\begin{equation}
\vec{F}=-Mg\hat{j}+\rho_{f} Vg\hat{j}+\sum_{i}\vec{f}_i^c
\end{equation}
where $\hat{j}$ is the unit vector along the vertical.

\begin{equation}
\vec{T}= \sum_{i}(\vec{x_{i}}-\vec{x_{cm}})\times\vec{f}_i^c
\end{equation}
with respect to the template's center of mass $\vec{r}_{cm}$.

The equations of motion of the particle template are:

\begin{equation}
\dot{\vec{U}}=\frac{\vec{F}}{M}
\end{equation}

and

\begin{equation}
\dot{\vec{\Omega}}=\frac{\vec{T}}{I}
\end{equation}

\noindent where $M$ is the mass of the template particle; $\vec{U}$ and $\vec{\Omega}$ are 
the linear and angular velocities of the template particle, respectively; $I$ is the 
moment of inertia; and $\vec{T}$ is the torque, \cite{Esa}, \cite{Kai}.

The boundary conditions at the container wall are zero for the normal velocity component 
of the fluid and no-slip condition for the tangential component, because the walls are 
assumed impenetrable, \cite{Kai}, \cite{Wachmann}.
The interaction between the oblate and the walls is defined through
a contact force, \cite{Esa}, where the walls are treated as a particle with infinite 
mass and infinite radius.  
\section{Results}
\begin{figure}
\begin{center}
\epsfig{file=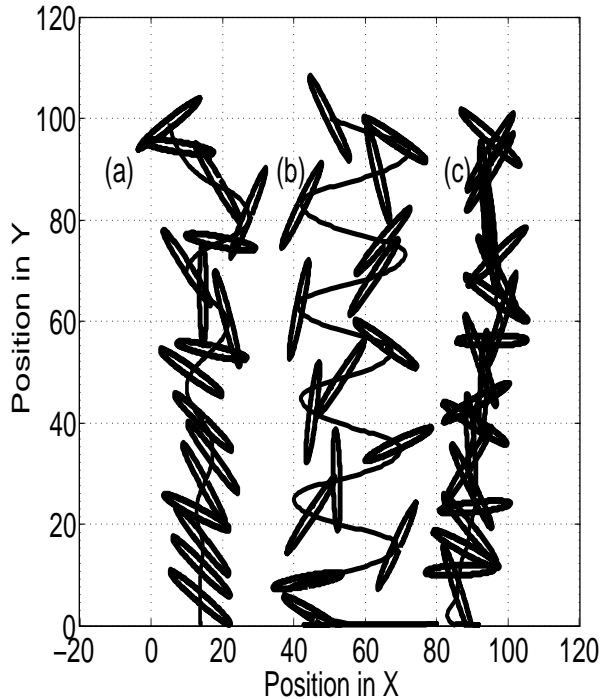,angle=0,width=8cm,height=9.5cm}
\vskip 1 cm        
\caption{(a) Steady-falling, (b) Periodic-oscillation, (c) Chaotic motion.} 
\end{center}
    
\end{figure}
For all our simulations, and in order to reduce the parameter space we use 
$\rho_{fluid}=1\frac{g}{cm^{3}}$ and $\rho_{oblate}=3.5\frac{g}{cm^{3}}$.

We found three basically different motions in our simulations: steady-falling, 
side-to-side periodic-oscillation known as 'flutter' ref.\cite{Belmonte}, and a 
chaotic motion as shown in fig. 2. The above phenomenology can be compared 
to the work of ref.\cite{Nori}, for the case of dropping disks. In general, the 
trajectories depend strongly on the initial conditions and the properties of the 
system (oblate's orientation $\Theta_o$, dynamical viscosity $\mu$ and the oblate's 
aspect-ratio $\Delta r$, etc). We don't find the tumbling motion described in 
the above references.

\begin{figure}
 \begin{center}

 \epsfig{file=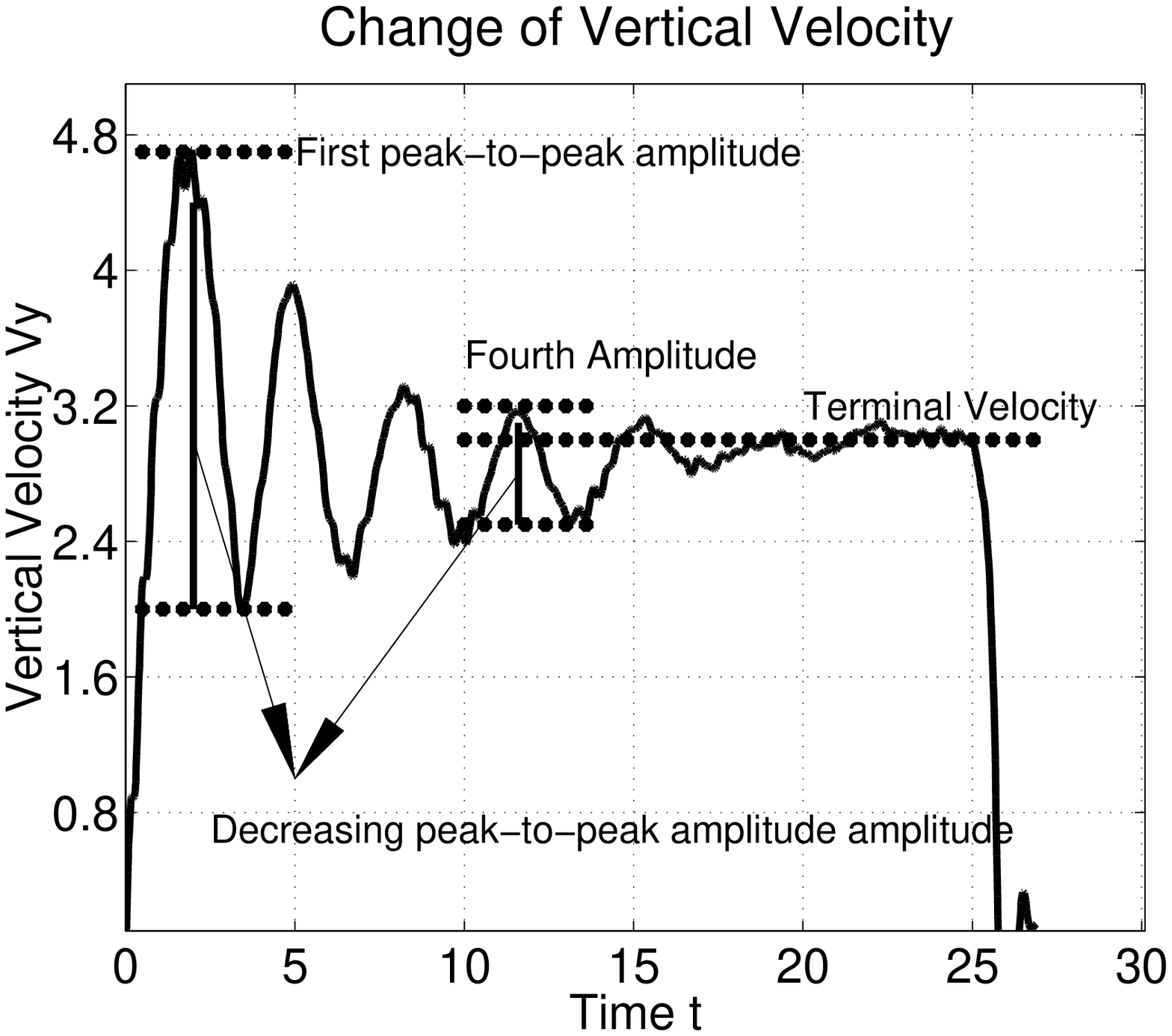,angle=0,width=7cm,height=6cm}

 \vskip 1 cm
 \epsfig{file=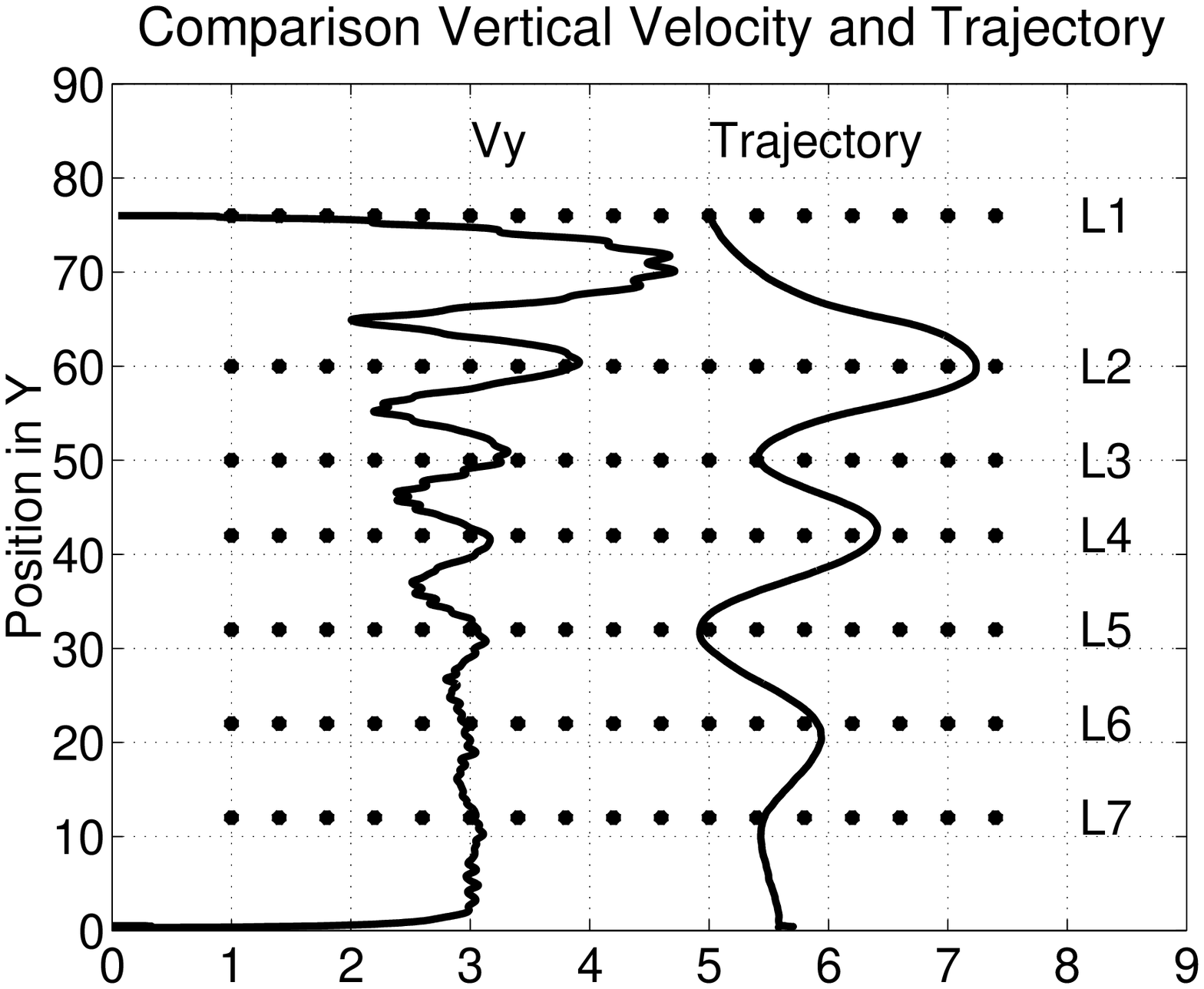,angle=0,width=7cm,height=6cm}
     
 \end{center} 
 \caption{(top) Decreasing amplitude of the vertical velocity. (bottom) Comparison 
      between the vertical velocity and the spatial trajectory at the same height.
  Initial conditions of the system. $\theta_o = 26.6^{0}$, $\Delta r = 0.25$, 
  $\mu = 0.033$. Falling initial height $h_{o}=76cm$ for the case of steady-falling.}  
  
\end{figure}

\subsection{Phenomenology of the Steady-Falling Oblate}

For the velocity of the center of mass, the vertical component  decreases when 
the oblate approaches the container bottom and shows a damped wavering, with an 
amplitude that decreases with time (fig. 3 top). 

The vertical trajectory is composed of succesive turning points, that correspond to 
the points where the trajectory changes the sign of the rate of change of the vertical 
velocity component marked in fig. 3(bottom) by the horizontal lines L2, ... L5. We 
also see, that from $L_1$ to $L_6$ the amplitude for the trajectory and the 
velocity decrease. 


\begin{figure}
 \begin{center}

 \epsfig{file=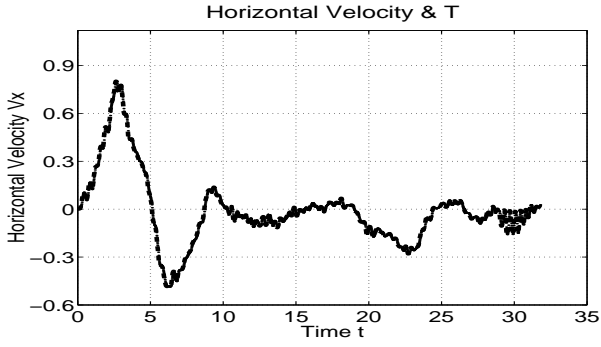,angle=0,width=8cm,height=4.5cm}

 \end{center} 
 \caption{Horizontal velocity. Initial conditions of the system. $\theta_o = 26.6^{0}$, 
$\Delta r = 0.25$, $\mu = 0.033$. Falling initial height $(b) h_{o}=76cm$.}  
  
\end{figure}

The horizontal components of the velocity obey a different behavior as the vertical ones, 
and in general they have non-regular oscillations as seen in fig.4.


\begin{figure}
 \begin{center}

 \epsfig{file=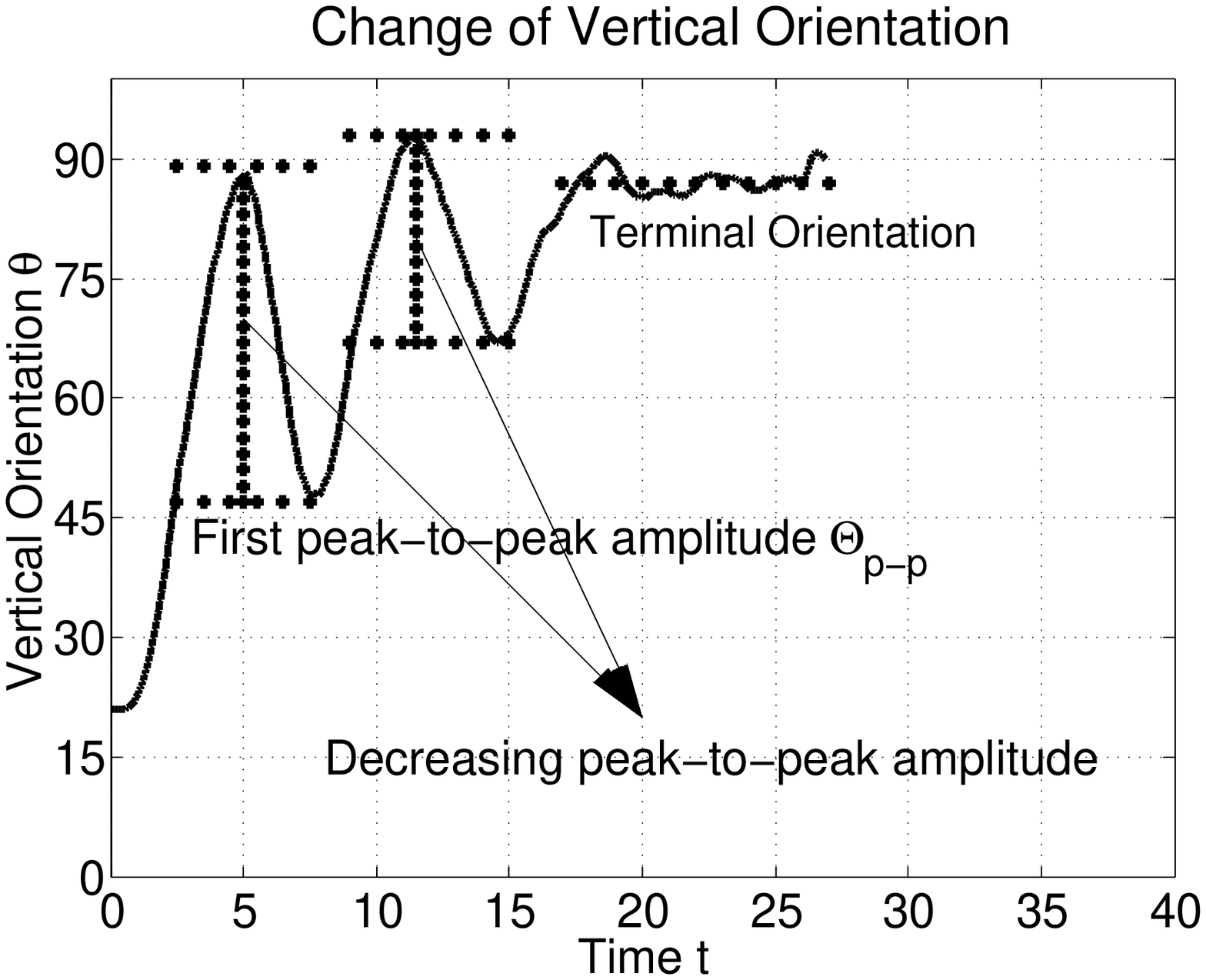,angle=0,width=7cm,height=6cm}

 \epsfig{file=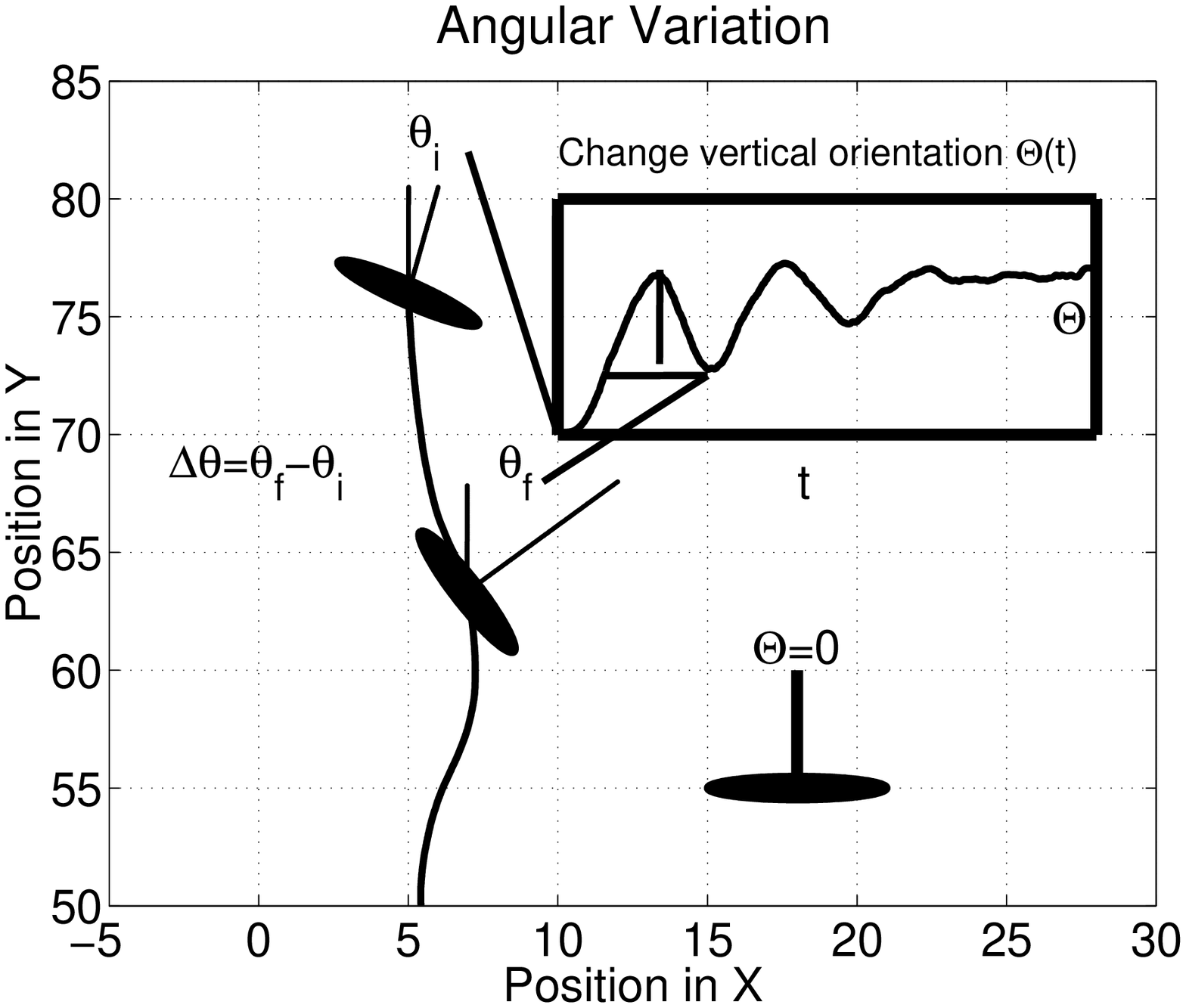,angle=0,width=7cm,height=7cm}

 \end{center} 
 \caption{(Top) Decreasing peak-to-peak amplitude in the angular oscillation. (Bottom) 
Angular change $\Delta \Theta$ along the vertical trajectory.  Initial conditions of the 
system. $\theta_o = 26.6^{0}$, $\Delta r = 0.25$, $\mu = 0.033$. Falling initial height $(b) h_{o}=76cm$.}  
  
\end{figure}

The oblate orientation is described through the three rotational degrees of 
freedom, around the center of mass. We present the time evolution of the 
angle between the oblate's normal and the vertical direction that we will call 
vertical orientation fig. 5 (top). $\Theta = 0$ implies that the oblate's principal axis 
will be parallel to the container's horizontal fig. 5 (bottom). At the beginning of the movement, 
there is a larger angular change of the oblate's normal $\Delta\Theta$ fig. 5(bottom), which 
is characterized by a larger peak-to-peak amplitude $\Theta_{p-p}$. In fig. 5 the definitions 
we illustrate the definition of the peak-to-peak amplitude as the distance between succesive
turning points, which decreases while the oblate sinks. In the steady-falling regime, both the 
peak-to-peak amplitude $\Theta_{p-p}$ and the change $\Delta\Theta=\Theta_{f}-\Theta_{i}$ , 
become smaller from the top towards the container's bottom. The oblate tends to align its 
major axis along the vertical \cite{Huang}, presenting the lowest resistance to its descent 
in the fluid, and acquiring a limit vertical velocity fig. 3 (top).


\begin{figure}
\begin{center}

\epsfig{file=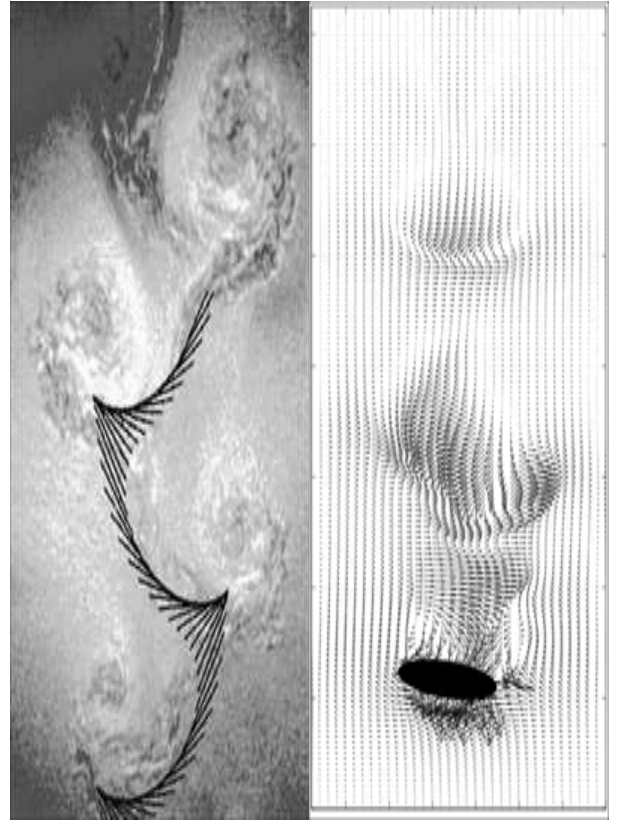,angle=0,width=8cm,height=11cm}
\vskip 0.5 cm        
 \caption{The right picture shows the vortex structure of the vertical and horizontal components 
 of the fluid velocity field $(u,v)$ generated by the falling oblate, with diameter of 3.2 cm in 
 a container of $10 \times 30 \times 30$ cm and Reynolds number of $Re=128$, aspect-ratio 
 $\Delta r=0.5$. The left picture shows shedding vortices reported by Belmonte et al, 
 ref.[12], for a falling strip.} 
 \end{center}
     
 \end{figure}
The oblate generates shedding vortices in the fluid along its falling trajectory, an example 
is shown in figure 6 (right), that shows the velocity fluid field around the oblate, and the 
vortex is localized just in the top region above the oblate, and where the oblate has experienced 
the larger angular change $\Delta\Theta$ as shown in fig. 5 (top). The Reynolds 
number calculated from the oblate's diameter and terminal velocity is $Re=128$. We point
out that the vortex structure is obtained also in the work of Belmonte et al, 
\cite{Belmonte} where a shedding vortex created by the zigzag motion of the falling strip 
is seen in the left figure 6. 
\vskip 1 cm
\subsection{Steady-Falling Oblate: Change in the Initial Height.}
\begin{figure}
 \begin{center}

\epsfig{file=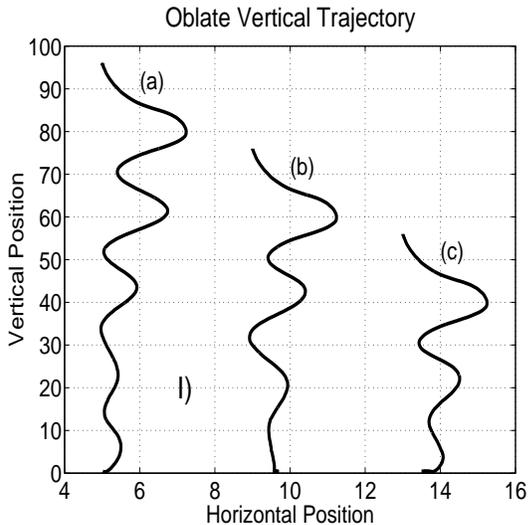,angle=0,width=7cm,height=7cm}

 \vskip 1 cm

\epsfig{file=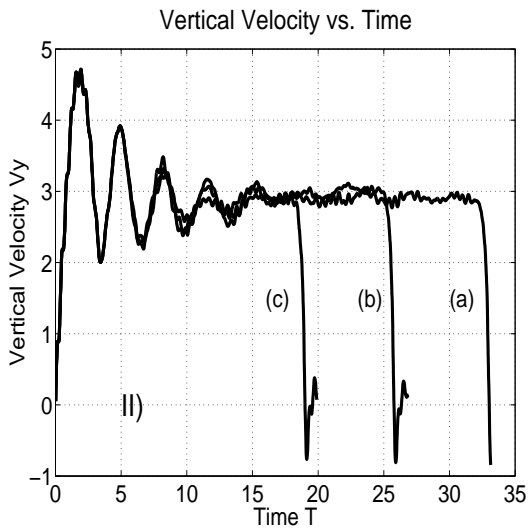,angle=0,width=7cm,height=7cm}

 \vskip 1 cm     

\epsfig{file=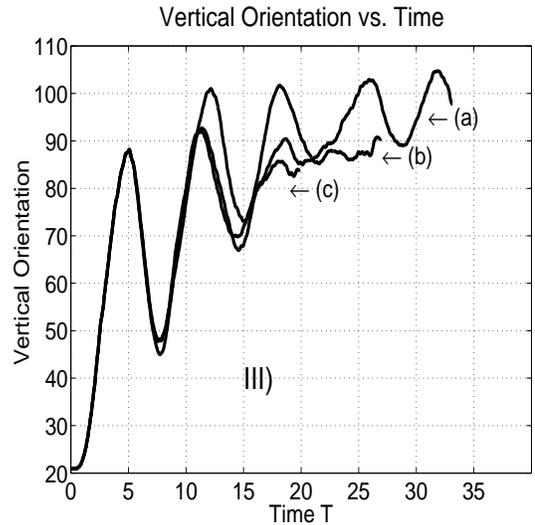,angle=0,width=7cm,height=7cm}

 \end{center} 
 \caption{ Initial conditions of the system. $\theta_o = 26.6^{0}$, $\Delta r = 0.25$, 
  $\mu = 0.033$. Each Trajectory has different initial height $(a) h_{o}=96cm$, 
  $(b) h_{o}=76cm$, $(c) h_{o}=56cm$. I) The spatial trajectory in the vertical plane. 
  II) Vertical velocity vs. time. III) Vertical orientation vs. time.}  
  
\end{figure}

We choose three different initial heights fig. 7I, (fixing the rest 
of parameters). If we superpose the trajectories 
in fig. 7I, it can be shown that there is no variation in the wavelength 
or in the peak-to-peak amplitude. For all the heights, the trajectories 
generated are in good agreement with a damped harmonic oscillation.

For the range of falling heights used, the oblate attains the same terminal 
vertical velocity fig. 7II, ($3 \frac{cm}{s}$), its magnitude is independent 
on the falling height. The vertical velocity finally, reaches a stable state 
(uniform and linear motion), after the same time ($\sim20 s$ fig 7II). We 
also see that the vertical velocity suddenly becomes zero when the oblate
touches the bottom.  

We see in fig. 7 III(c-a), an increase in the final angle with respect to
the initial height. For $h_{o}=56$ the smaller height, we obtain $\sim85^{o}$ and 
for larger one $h_{o}=96$, we have $\sim95^{o}$. For the larger height the oblate 
is still in a transitory state before it arrives to its final angle fig. 7III.

\subsection{ Steady-falling oblate: Change in the dynamical viscosity.}
\begin{figure}
\begin{center}

\epsfig{file=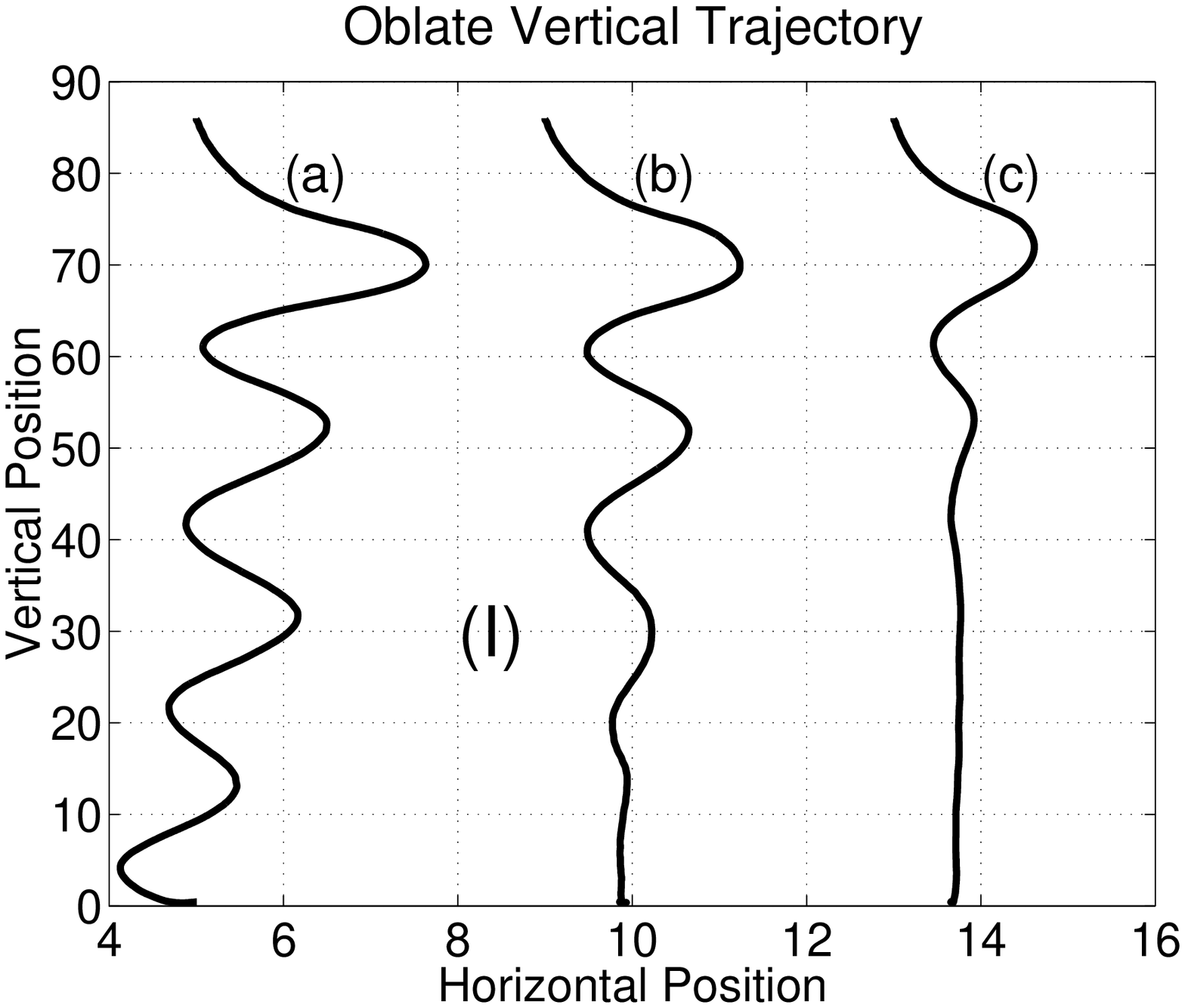,angle=0,width=7cm,height=7cm}

\epsfig{file=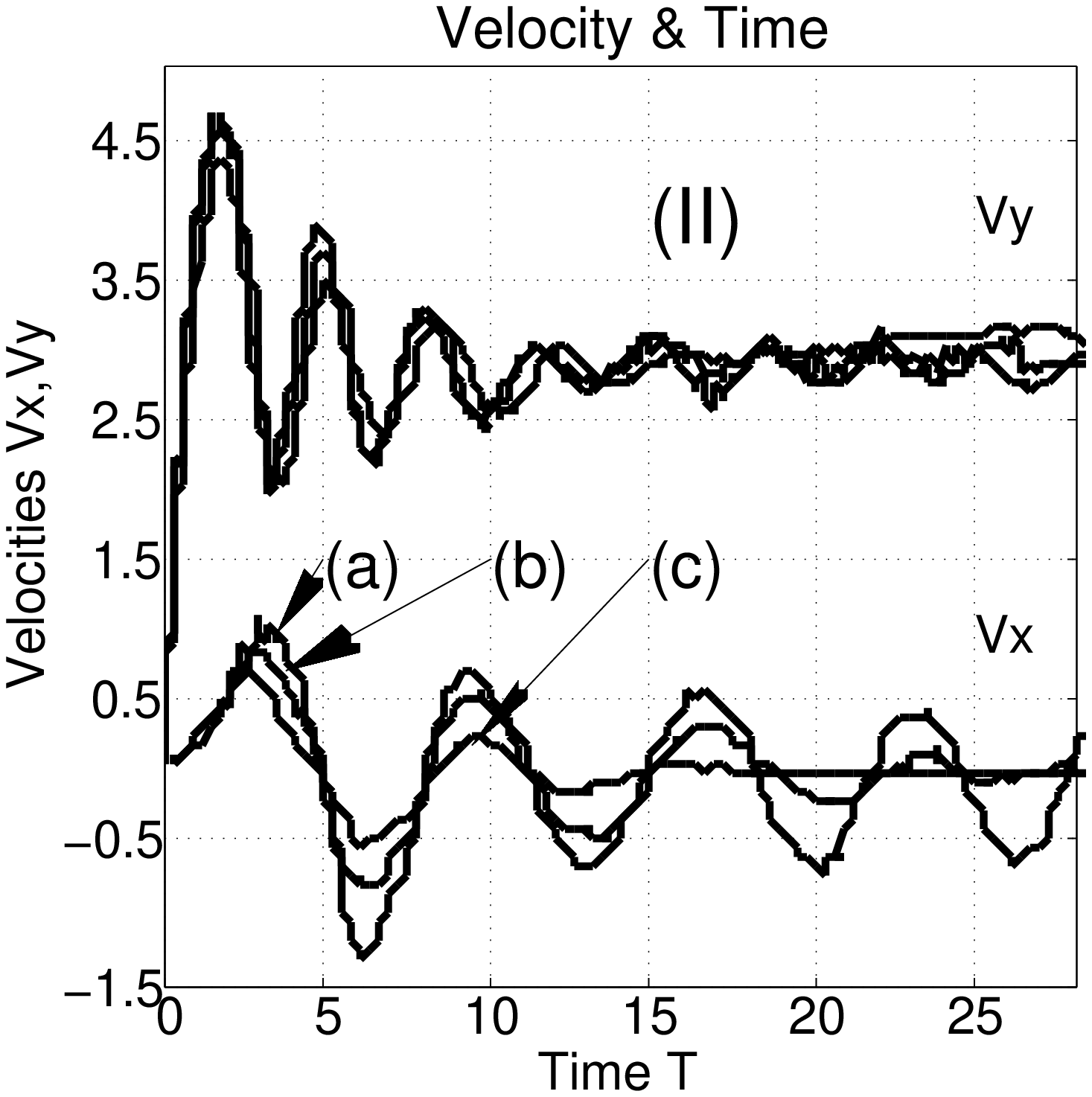,angle=0,width=7cm,height=7cm}

\epsfig{file=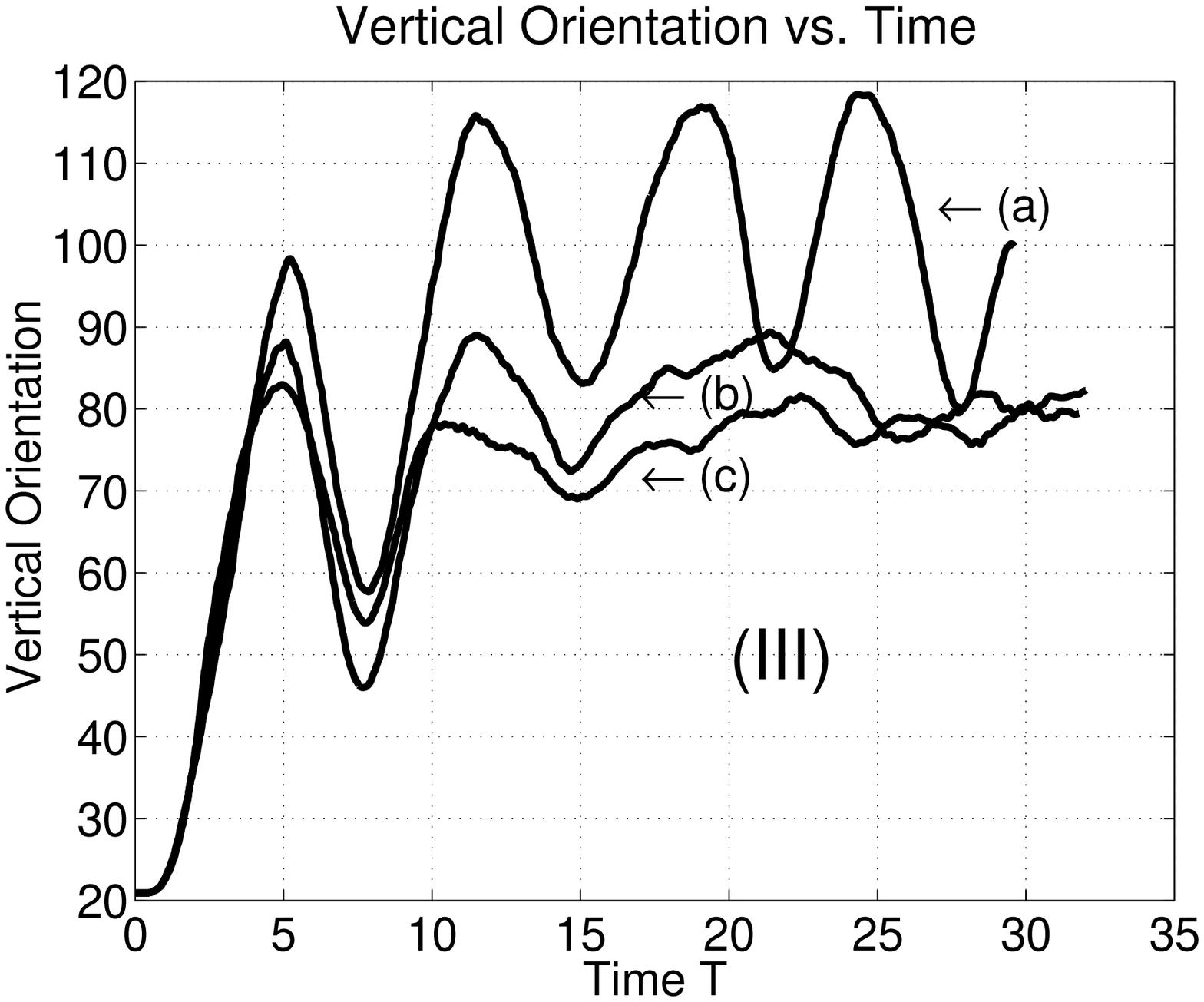,angle=0,width=7cm,height=7cm}

\end{center}

\caption{Initial conditions in the system. $\theta_o = 26.6^{0}$, $ h_o = 80 $, 
  $\Delta r = 0.25$. Each trajectory has a different dynamical viscosity $(a) \mu=0.025 $, 
  $(b) \mu=0.033$, $(c) \mu=0.100$. I) Proyection of the trajectory in the vertical plane. 
  II) Vertical (top) and horizontal (bottom) velocities vs. time. III) Vertical orientation 
  vs. time .}  
   
\end{figure} 

The oblate starts its swinging motion with a given initial orientation $\theta_0=26.6^{o}$. 
It glides downwards and to the side acquiring some amplitude, while the dynamical 
viscosity $\mu$ acts reducing the subsequent amplitudes of oscillation (fig. 8I). 
As we increase the dynamical viscosity from $(a) \mu=0.025 $ to 
$(c) \mu=0.100$ the attenuation in the oscillatory trajectory becomes stronger. 
In fig. 8I a, there is a weak damping producing a long oscillatory behavior and 
in fig. 8I c, the trajectory is quickly attenuated in the first half of the 
falling height, and hereafter it follows a vertical trajectory. Similar 
decreasing oscillations, are observed when one small sphere 
suspended from a fixed point by a string or a rod, oscillates against the air 
and generates a damped harmonic oscillator.
  
The behavior of the vertical velocity is shown in fig. 8II. There is a clear 
damping in the peak-to-peak velocity amplitude, in all the cases. The attenuation 
in the vertical velocity amplitude and the time between two consecutive turning 
points are not very different from each other, as the fluid dynamical viscosity 
is changed. For the three viscosities the oblate adquires approximately the same 
final mean velocity $v_y=3\frac{cm}{s}$. 

A larger variation in the velocity is observed in its horizontal component fig. 8II. 
There is a strong attenuation in the curve for larger viscosities fig. 8IIc. One 
explanation is that the interaction between the walls and the oblate is smaller
when the viscosity is decreased \cite{Brenner}. This is also supported by the 
larger angular variation $\Delta \Theta$ in fig. 8IIc, for smaller viscosities.

In fig. 8 III we see that the first peak-to-peak amplitude are nearly the same for all
considered viscosities. For $\mu=0.025$ the subsequent peak-to-peak amplitudes attain 
a constant value of $\Theta_{p-p}\sim 15^{o}$ while for higher viscosities periodic 
oscillations are replaced by erratic motion of low amplitude.

\subsection{Steady-Falling Oblate: Change in the Aspect-Ratio of the Oblate.}
\begin{figure}
 \begin{center}

  \epsfig{file=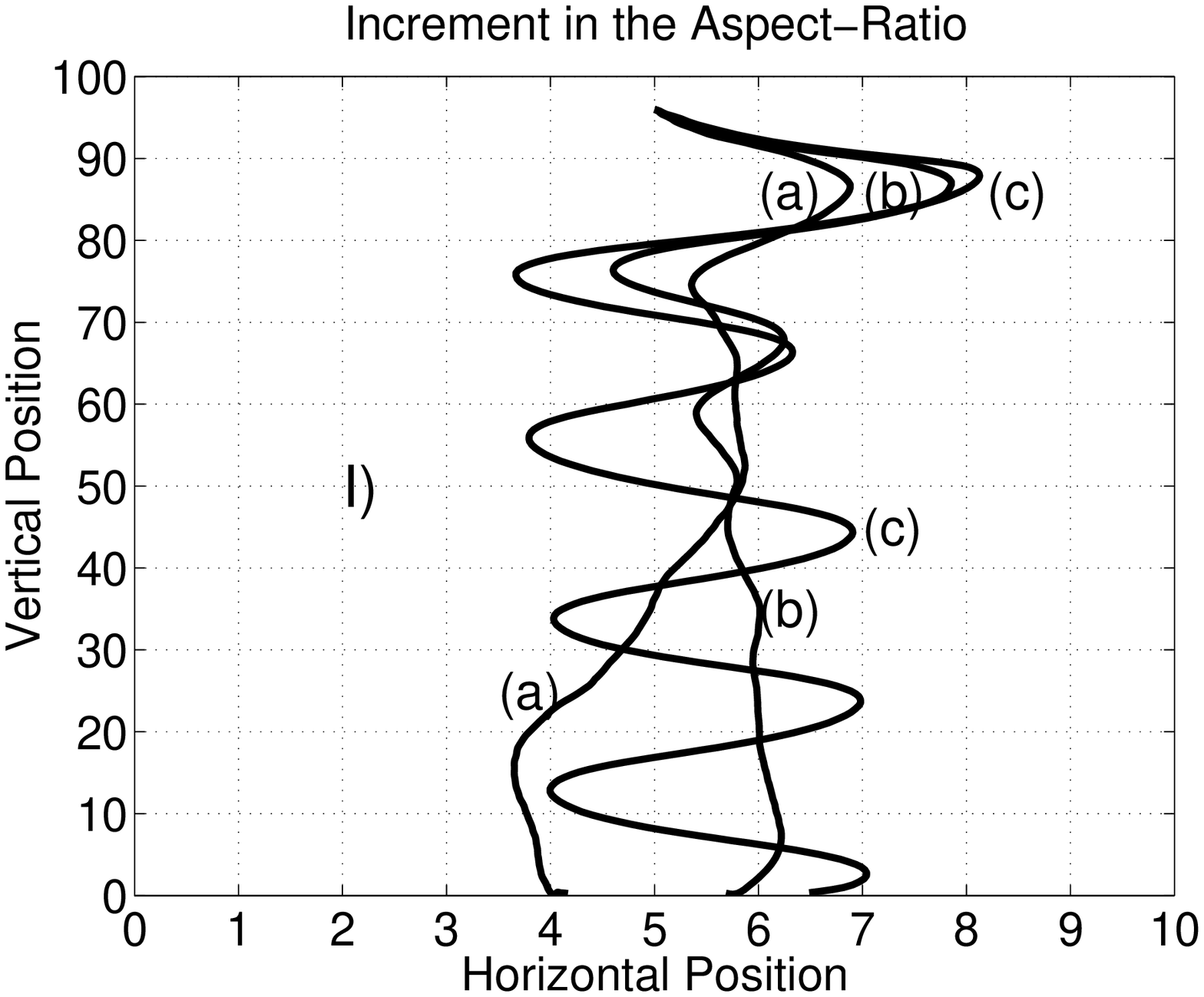,angle=0,width=7cm,height=7cm}

  \epsfig{file=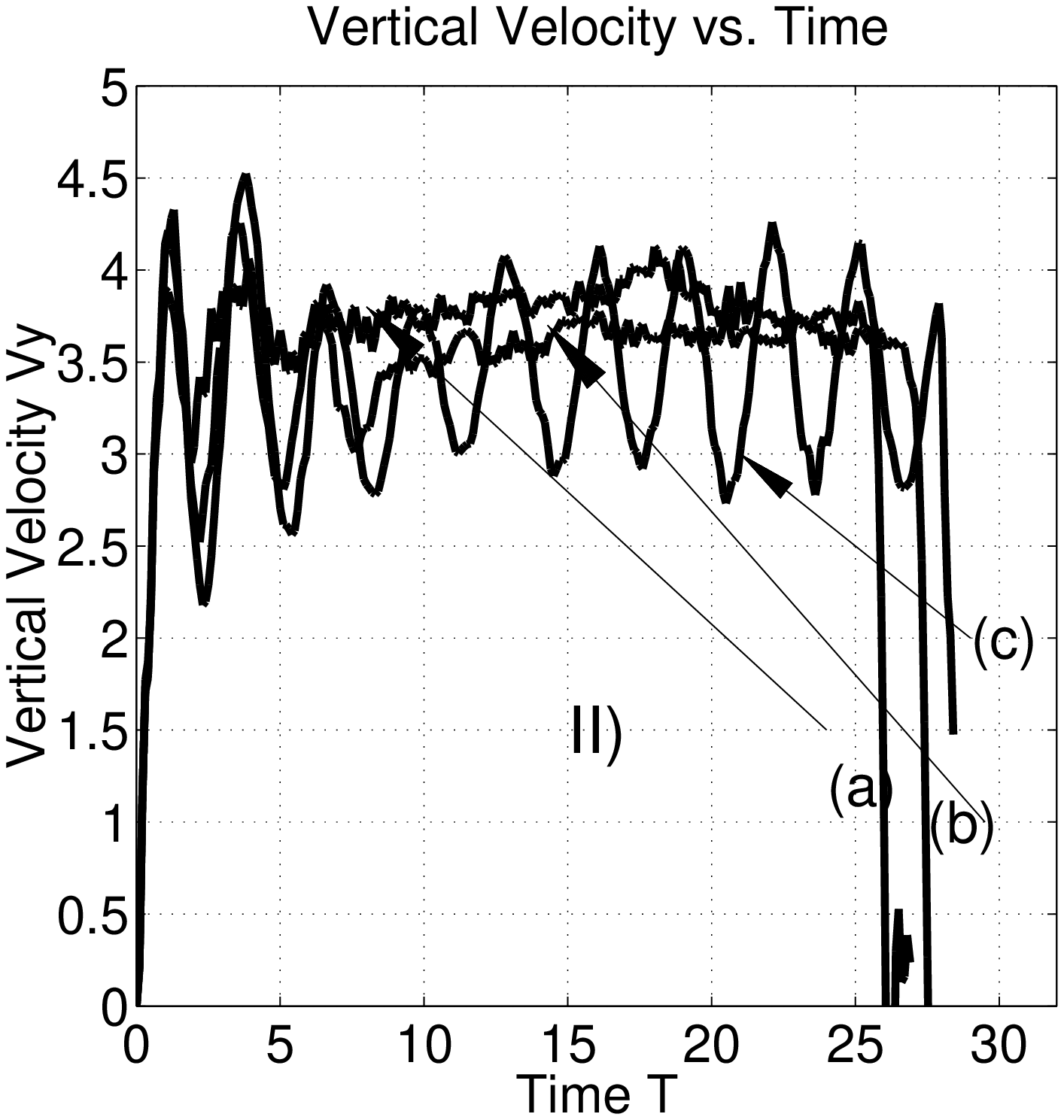,angle=0,width=7cm,height=7cm}

 \epsfig{file=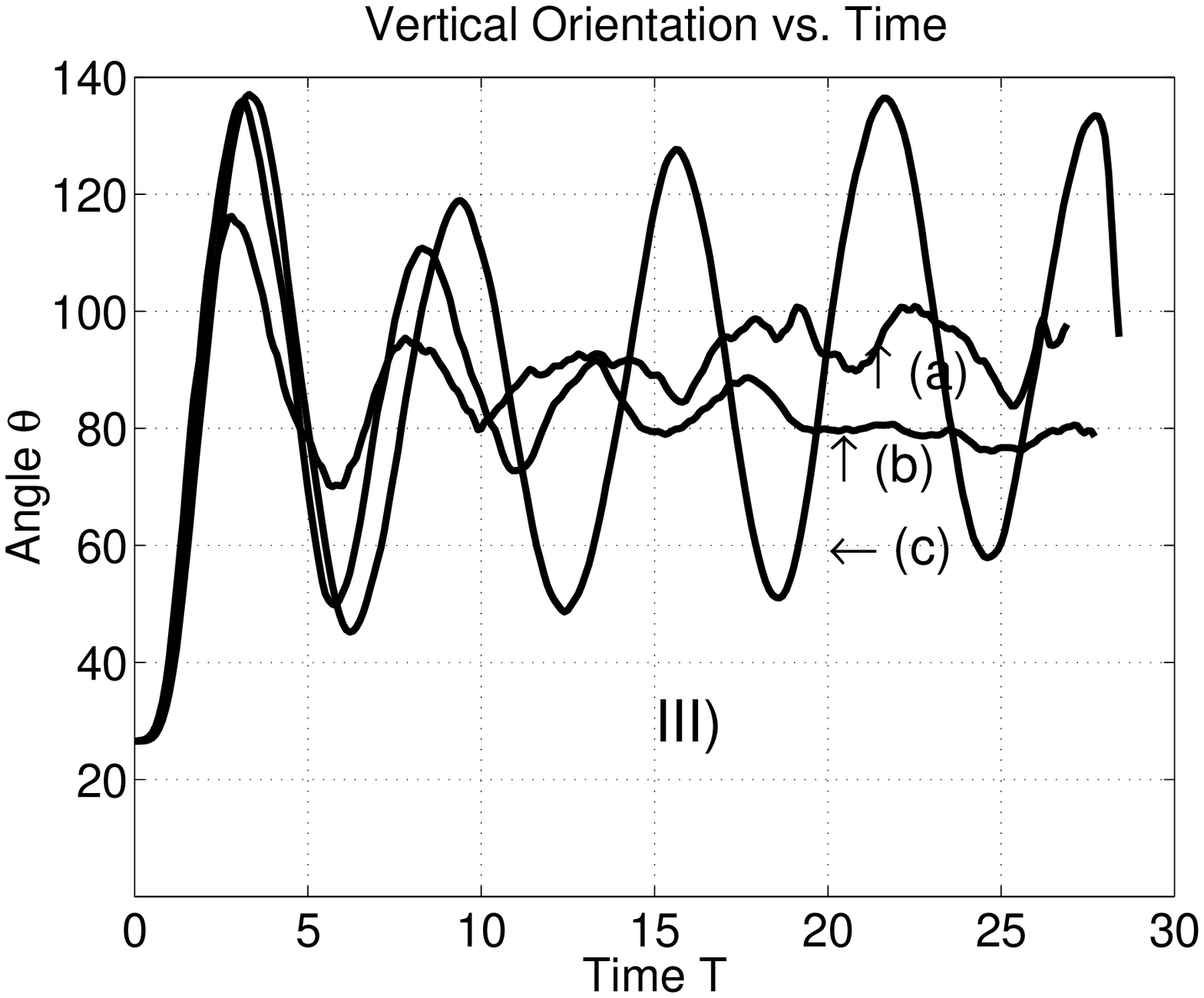,angle=0,width=7cm,height=7cm}
    
  \end{center} 
  \caption{Initial conditions of the system. $\Theta_o = 26.6^{0}$, $ h_o = 96 $, 
    $L_{x} \times L_{y} \times L_{z} = 10\times 100 \times 10$, $ \mu=0.033$. Each 
   trajectory has different aspect-ratio $ (a) \Delta r=0,29$, $(b) \Delta r=0,22$, 
   $(c) \Delta r=0,18 $. I) Trajectory for the vertical plane. II) Vertical 
   velocity vs. time. III) Vertical orientation vs. time.}   
   
\end{figure}

When the oblate's falling motion begins, the oblate gains a larger oscillation 
amplitude in its oscillatory falling trajectory. And for all the three trajectories 
presented in fig. 9, this first amplitude decreases $A_o=3.0, 2.8, 1.8 cm$ as 
the oblate's aspect-ratio is increased $\Delta r=0.18, 0.22, 0.29$ respectively. This 
first large oscillation is a common characteristic for these trajectories. As the 
aspect-ratio is incremented, the number of cycles and the amplitude of the 
trayectories change. For $\Delta r=0.22$ the trajectory presents a well defined 
steady-falling behavior, fig. 9I(b), but, if the aspect-ratio increases, 
($\Delta r=0,29$), the trajectory varies, the peak-to-peak amplitude is quickly 
damped in the first half of the vertical trajectory, with the interesting observation 
that in the second half the trajectory doesn't have a steady-falling behavior fig. 9 Ia. 
When $\Delta r=0.18$, the smaller aspect-ratio, the trajectory has a oscillatory 
behavior fig. 9I(c), with a constant peak-to-peak amplitude of $3 cm$.

Since the minor oblate radius is fixed in our simulations, when the aspect-ratio 
is increased $\Delta r=0.18, 0.22, 0.29$, the oblate's area gets smaller, and
the final vertical velocity increases $Vy=3.5, 3.7, 3.9 \frac{cm}{seg}$, respectively, 
fig. 9II. The final vertical velocity decreases with the decrement in the aspect-ratio, 
since the smaller aspect-ratio presents more area against the fluid.
As the aspect-ratio is increased the peak-to-peak amplitude in the vertical velocity 
diminishes. For the smaller aspect-ratio $\Delta r=0.18$, we have the larger 
amplitude $1\frac{cm}{s}$, and as the aspect-ratio is increased $\Delta r=0.22$ 
and $\Delta r=0.29$ the peak-to-peak amplitude tends to be much smaller.

The peak-to-peak amplitude for the vertical orientation $\Theta_{p-p}$ increases 
when the aspect-ratio decreases or the oblate's area increases. 
For $\Delta r=0.29$ the peak-to-peak amplitude is $\Theta_{p-p}=15^{o}$, (fig. 9a), 
and much smaller compared to $\Theta_{p-p}=70^{o}$, (fig. 8c), for 
$\Delta r=0.18$. In all cases the oblate at the end orients vertically (fig. 9 III). 

\subsection{Periodic Behavior of a Falling Oblate.}
\begin{figure}
 \begin{center}

 \epsfig{file=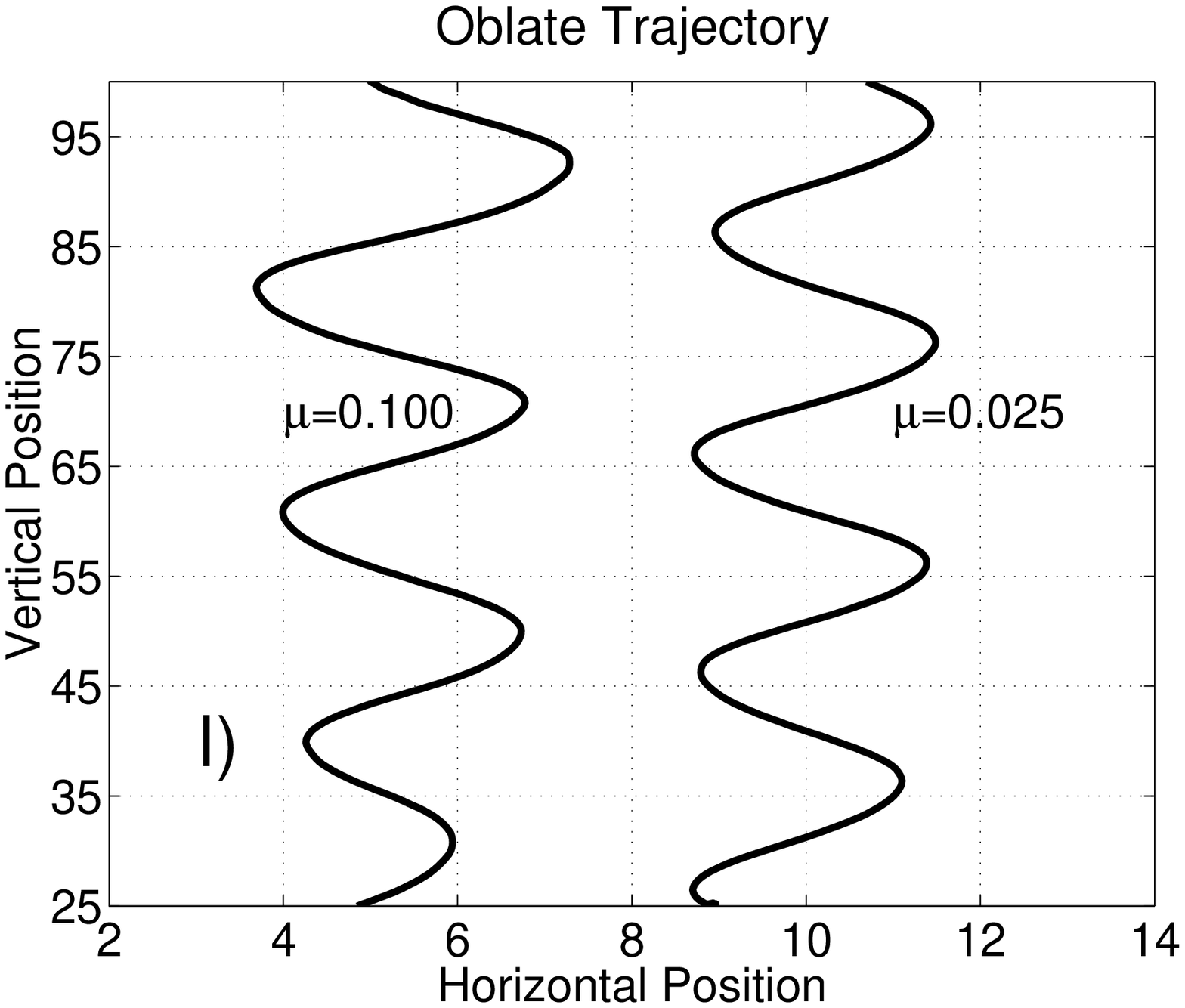,angle=0,width=7cm,height=6cm}

 \epsfig{file=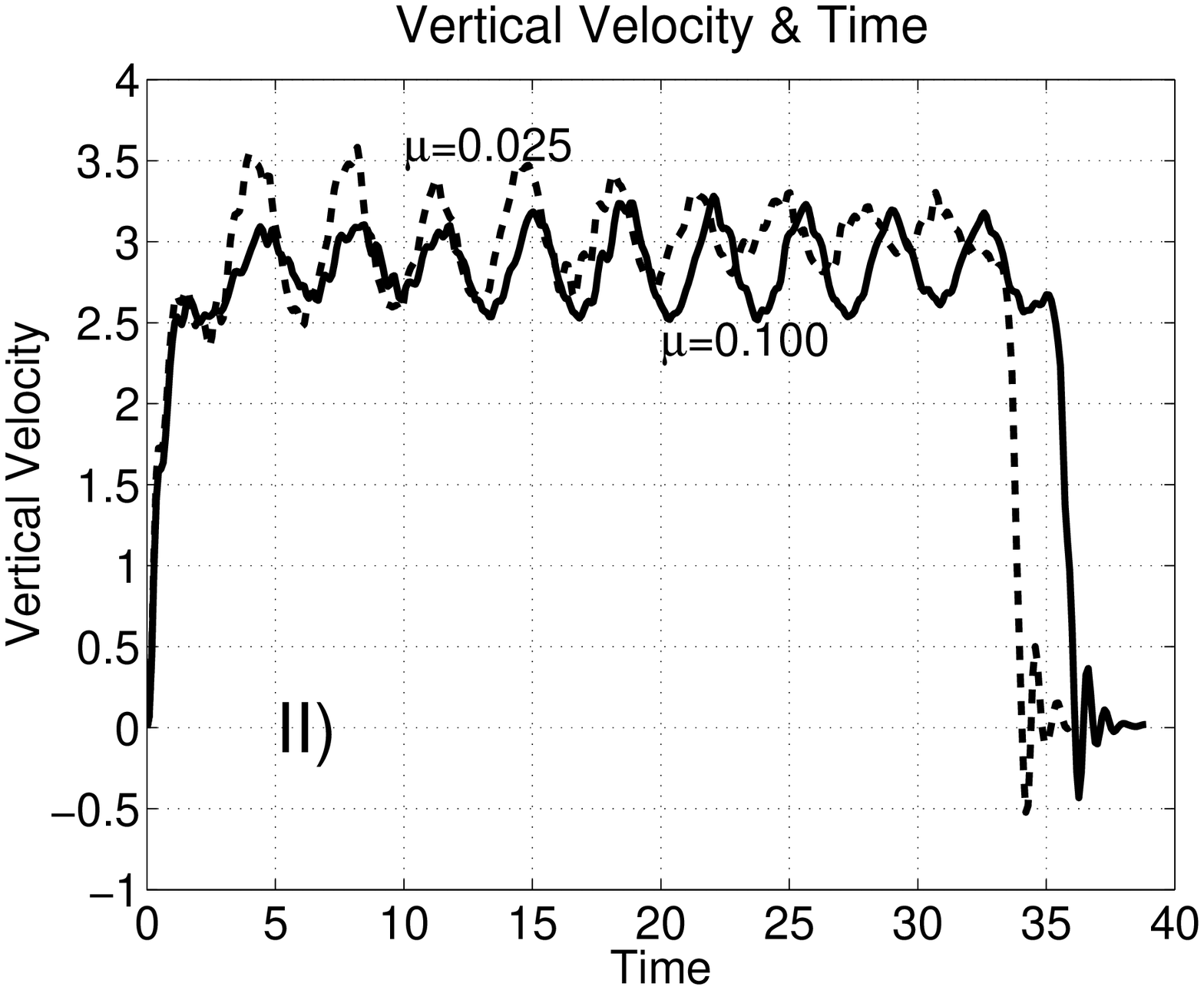,angle=0,width=7cm,height=6cm}

 \epsfig{file=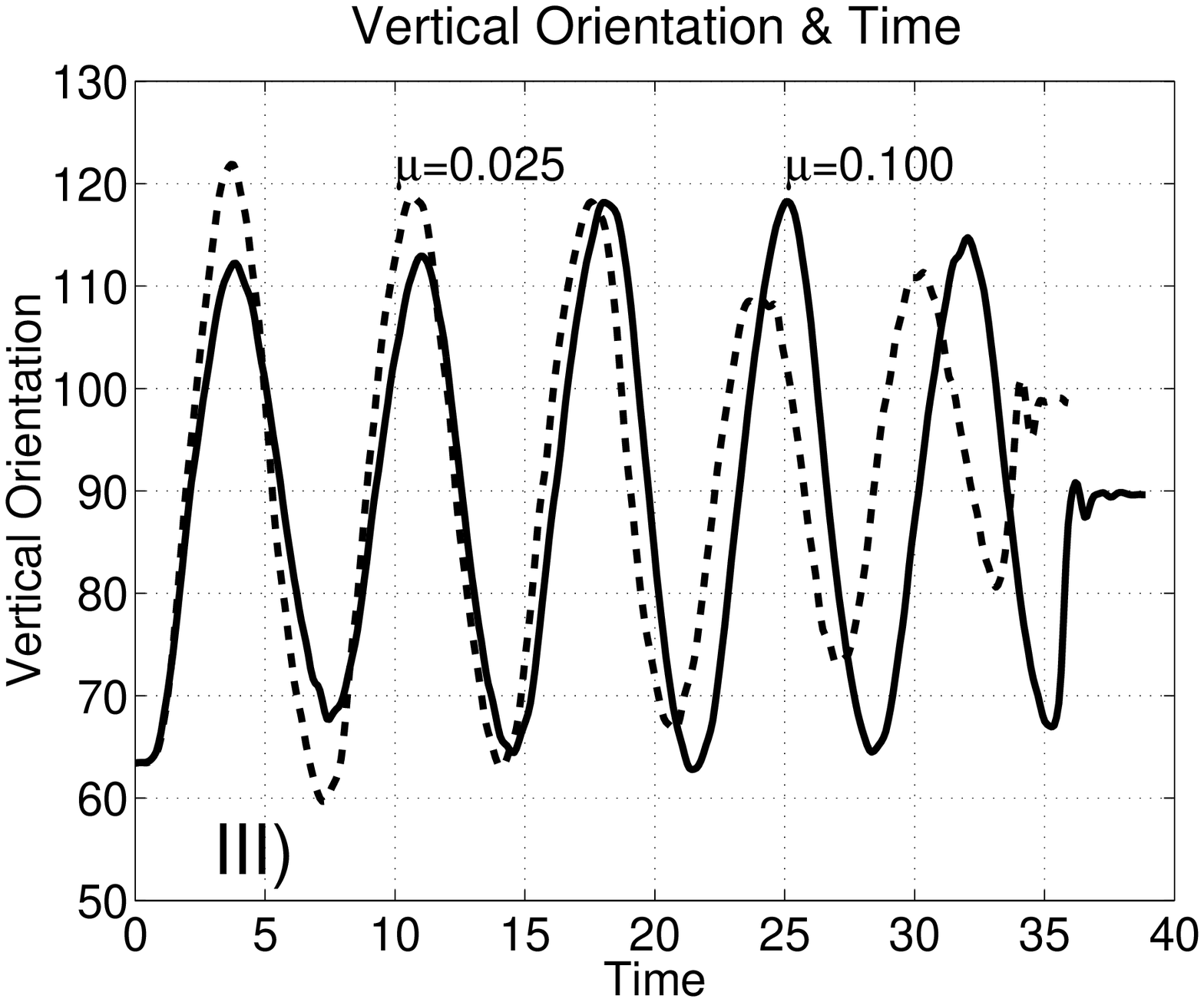,angle=0,width=7cm,height=6cm}

\end{center} 
 \caption{Trajectories generated for $\mu_{1} = 0.100$, $\mu_{2} = 0.025$. (I) Trajectory in 
 vertical and horizontal position. (II) Vertical orientation $\Theta$ vs time. (III) 
 Vertical velocity $v_y$ vs time. The initial conditions are $ h_o = 96 $, $\Delta r = 0.133$,
 $ \theta_0 = 63.4^{o} $.  }  
  
\end{figure}

We have found periodic behavior for smaller dynamical viscosity ($Re=480$) and smaller 
aspect-ratio ($\Delta r = 0.133$). The dynamics of the falling oblate is governed by 
inertial effects. In figure 10 (I), we show the transition from a quasi-periodic, or a 
long steady-falling trajectory($\mu_{1} = 0.100$), to a periodic behavior fig. 10
(I,$\mu_{2} = 0.025$), when the dynamical viscosity is varied from $\mu_1 = 0.1$ to 
$\mu_2 = 0.025$. The trajectory presented in fig. 10I, with dynamical viscosity 
$\mu_2 = 0.025$ has a wave length of 20 cm. 

The vertical velocity shown in fig. 10 (II), has the same transition from a long 
steady-falling regime with a final average velocity of $3.0\frac{cm}{s}$ to the 
periodic regime where the velocity has a oscillation  period of 3.3 s.

The vertical orientation presented in fig. 10 (III) has also the same transition from 
a long steady-falling regime to periodic behavior with a period of 6.6 s, and the 
angular values oscillate around $\Theta_{0}=90^{o}$ with angular peak-to-peak 
amplitude $\Theta_{p-p}=60^{o}$.

\vskip 1 cm
\begin{figure}
\begin{center}

\epsfig{file=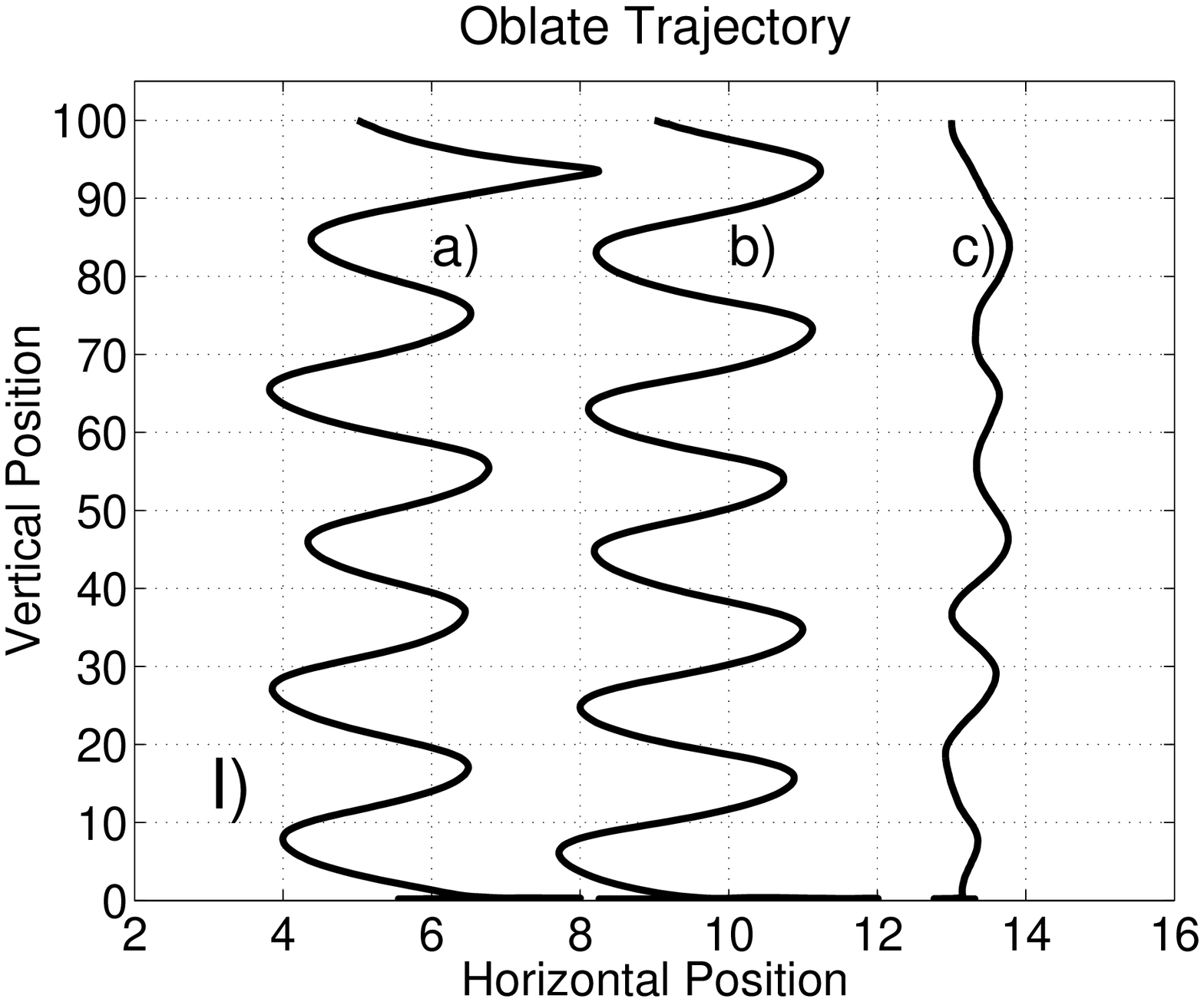,angle=0,width=7cm,height=7cm}

\vskip 1 cm

\epsfig{file=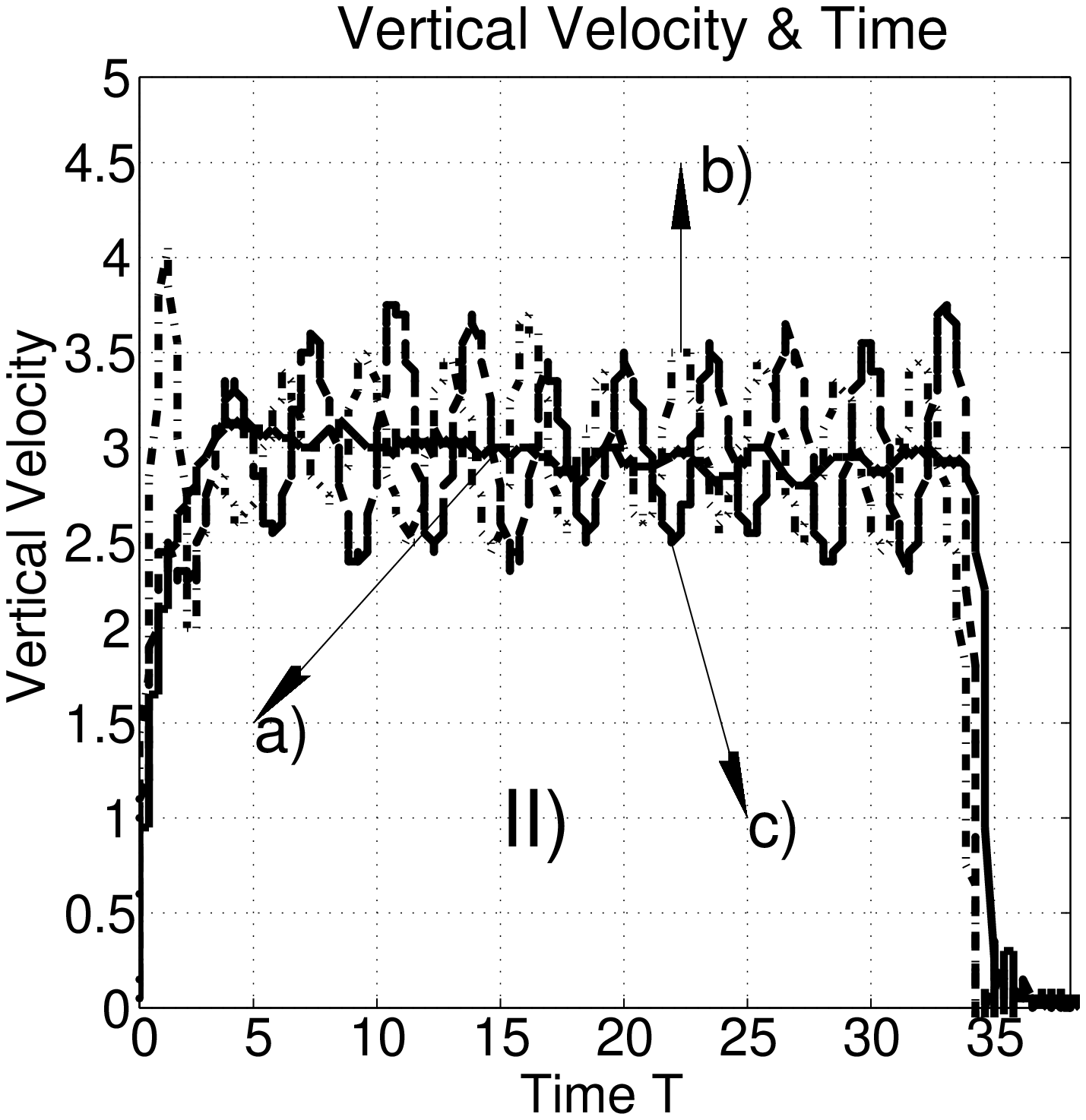,angle=0,width=8cm,height=7cm}

\vskip 1 cm

\epsfig{file=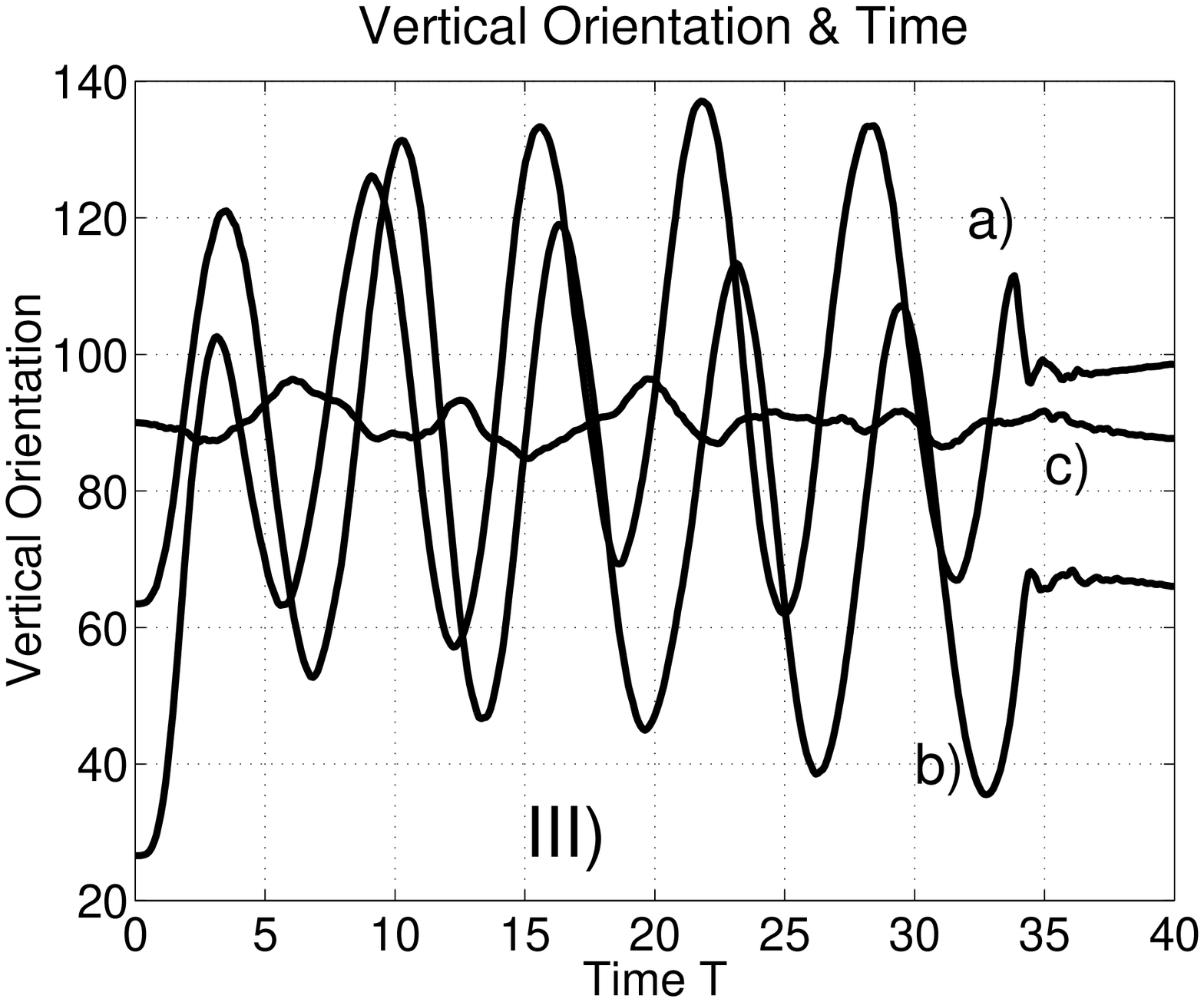,angle=0,width=7cm,height=7cm}

\end{center} 
\vskip 1 cm
\caption{I) Trajectories for three initial orientations. (a) $\Theta_0=26^{o}$ (b) 
 $\Theta_0=63^{o}$ (c) $\Theta_0=90^{o}$ . II) The corresponding vertical velocities. 
 III) The vertical orientations. Initial conditions $h_o = 100$, 
 $L_{x}\times L_{z}=10 \times 10$, $\Delta r = 0.133$, $Re=435$. } 
\vskip 0.5cm

\end{figure}
\vskip 1 cm
We perform three simulations in the periodic regime with very different initial 
orientation angles and the corresponding trajectories are shown in fig. 11I. In the 
case of $\Theta_0=26^{o}$ the peak-to-peak amplitude is 2.3 cm and for 
($\Theta_0=90^{o}$) it is 0.4cm, and the oscillatory behavior is observed for the 
three cases. The peak-to-peak amplitude of the oscillation in the trajectory fig.10I, 
decreases as $\Theta_o$ is increased. In the case of the vertical velocity and 
orientation fig. 11II-III, the initial orientation angle also plays the same role, 
reducing the peak-to-peak amplitude of the curves, and for $\Theta_o=90^{o}$ the 
amplitude of oscillation is the smallest of the three.

The final vertical velocity and orientation for the three trajectories are 
$3.0\frac{cm}{s}$ and $85^{o}$ respectively. We can say that the average final values 
for the vertical velocity and orientation are not modified by the variation of the initial 
orientation $\Theta_o$. 

We have found the largest peak-to-peak amplitude $1.0\frac{cm}{s}$ for the vertical 
velocity oscillation for an initial orientation of $\Theta_0=26^{o}$ and it 
becomes smaller as the oblate's initial orientation tends to $\Theta_0=90^{o}$. For 
the peak-to-peak amplitude of the vertical orientation the oblate shows a similar 
behavior: we have the larger value for the amplitude $\Theta_{p-p}=70^{o}$, and it 
is obtained for an initial orientation of $\Theta_0=26^{o}$. The smallest peak-to-peak 
amplitude $\Theta_{p-p}\sim4^{o}$ is obtained for $\Theta_0=90^{o}$.


\begin{figure}
\begin{center}
  \epsfig{file=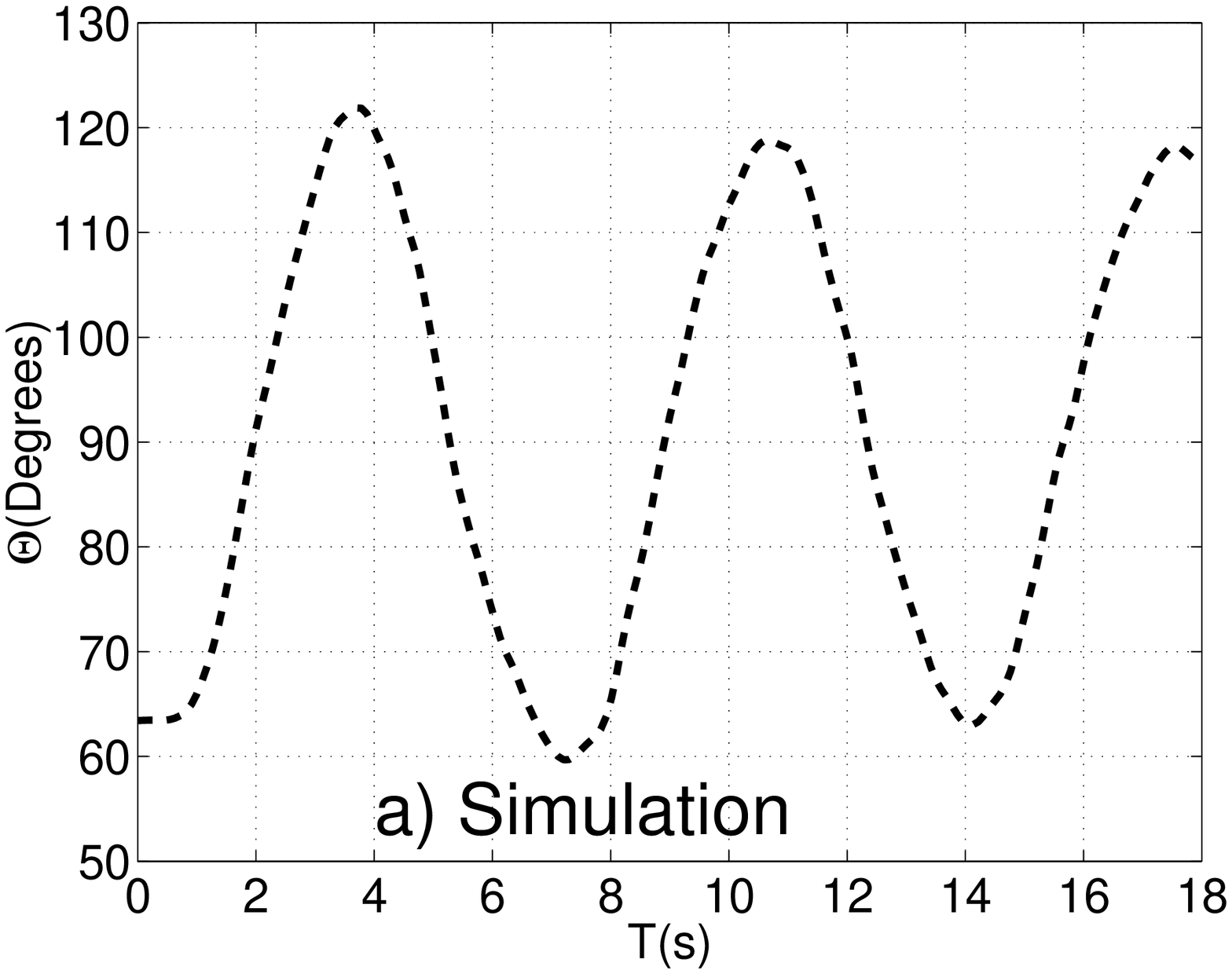,angle=0,width=7.5cm,height=5cm}
\vskip 0.5 cm
  \epsfig{file=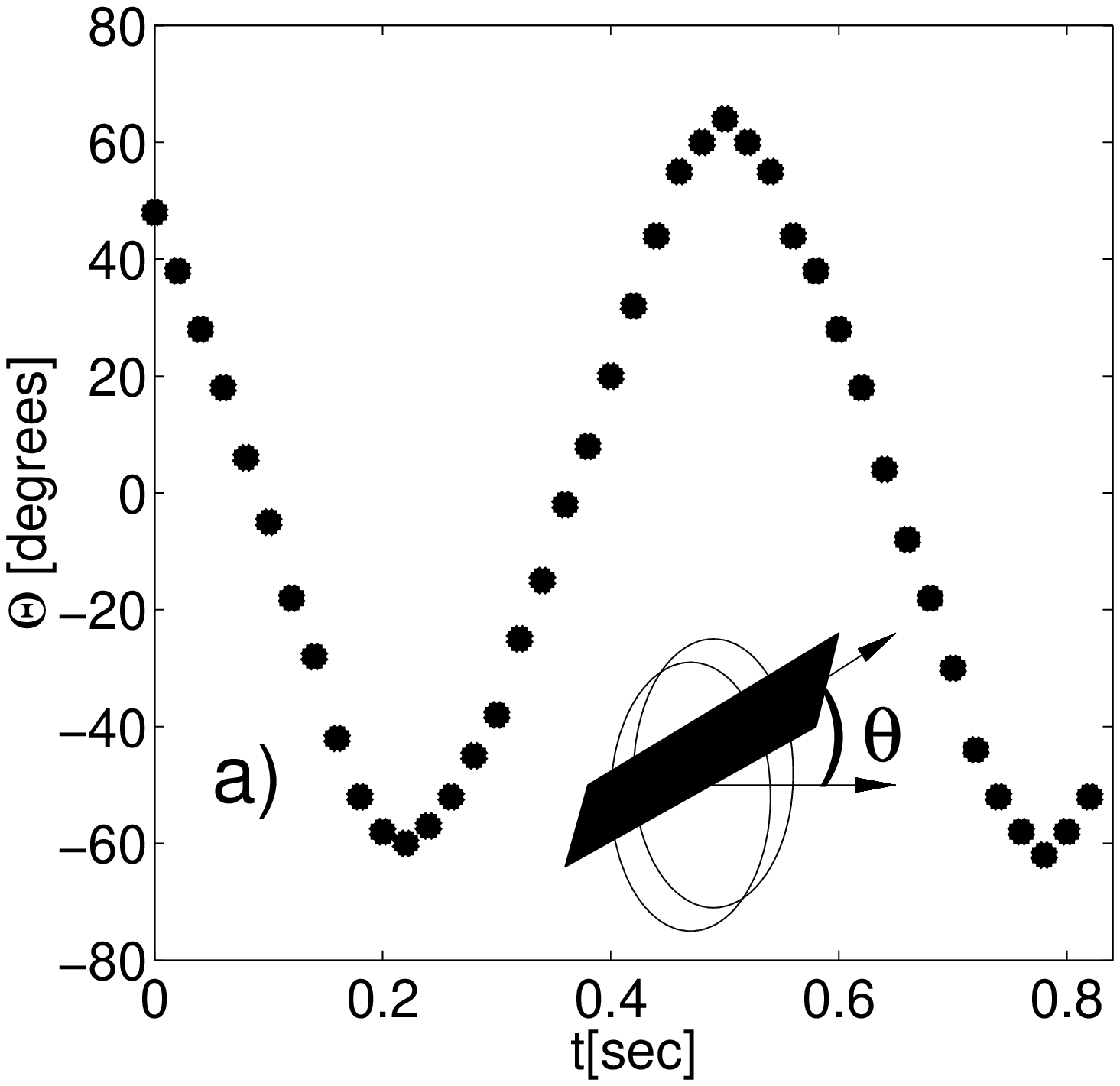,angle=0,width=7.5cm,height=5cm}
\vskip 0.5 cm
\epsfig{file=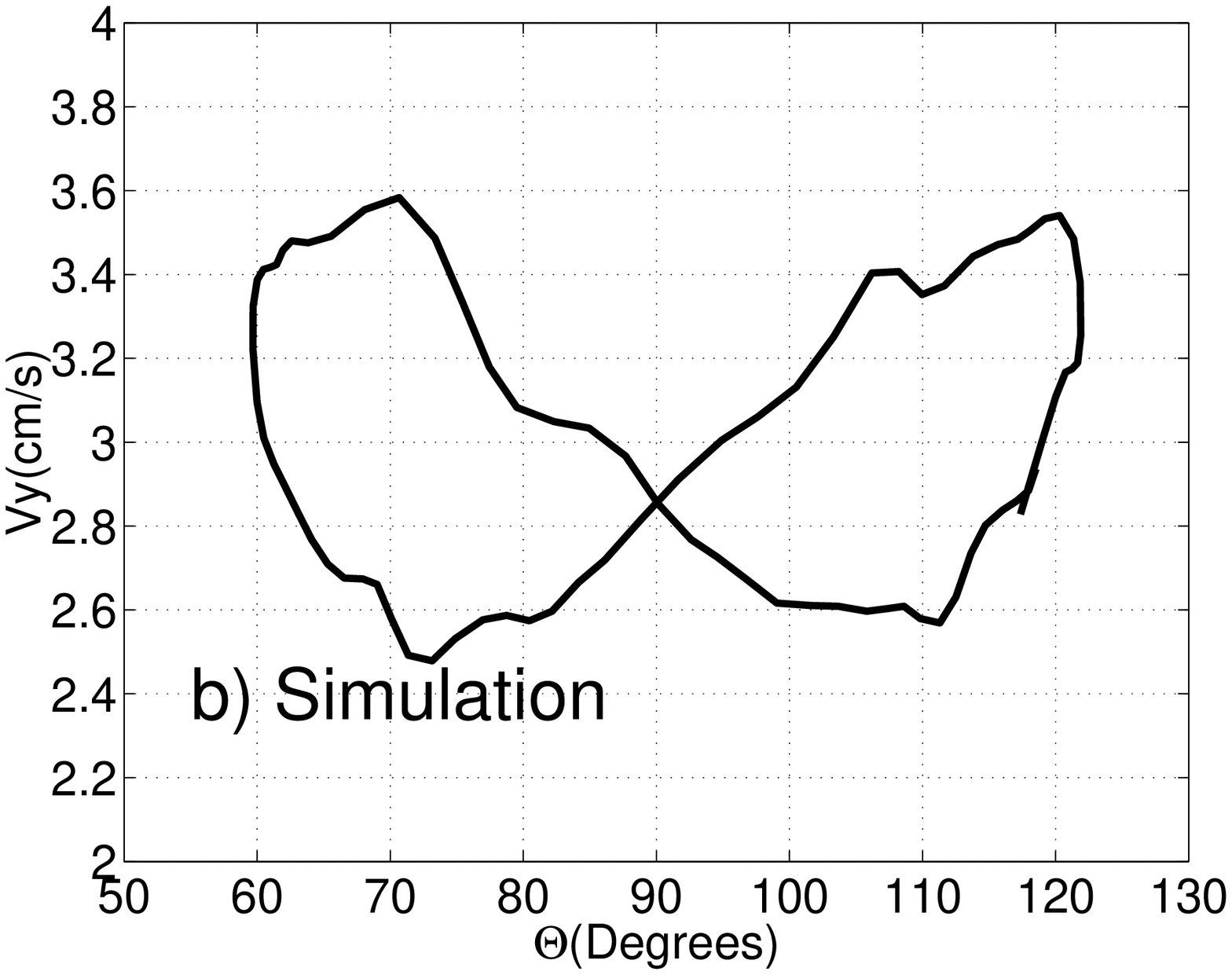,angle=0,width=7.5cm,height=5cm}
\vskip 0.5 cm
\epsfig{file=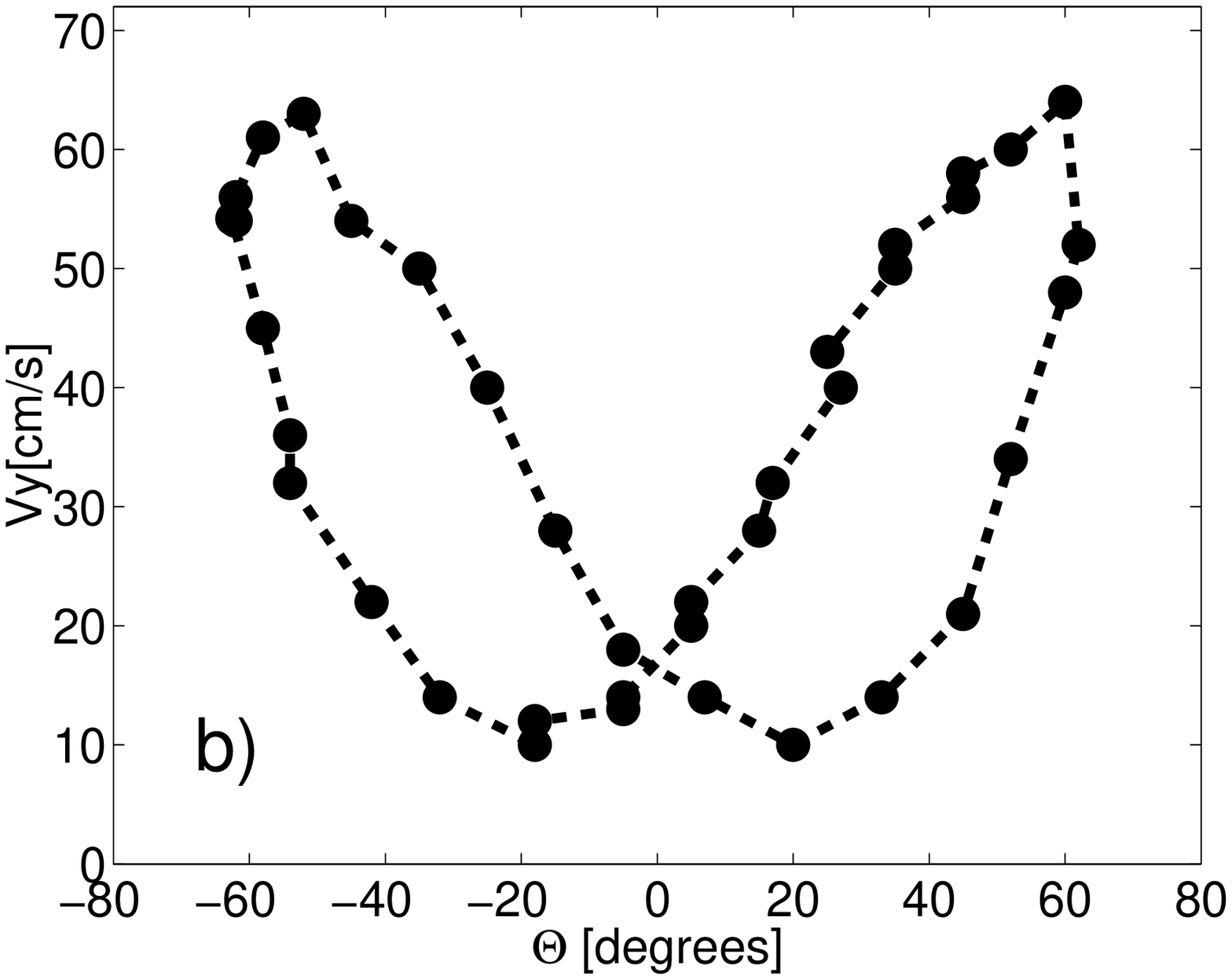,angle=0,width=7.5cm,height=5cm}
\vskip 1.0 cm
\epsfig{file=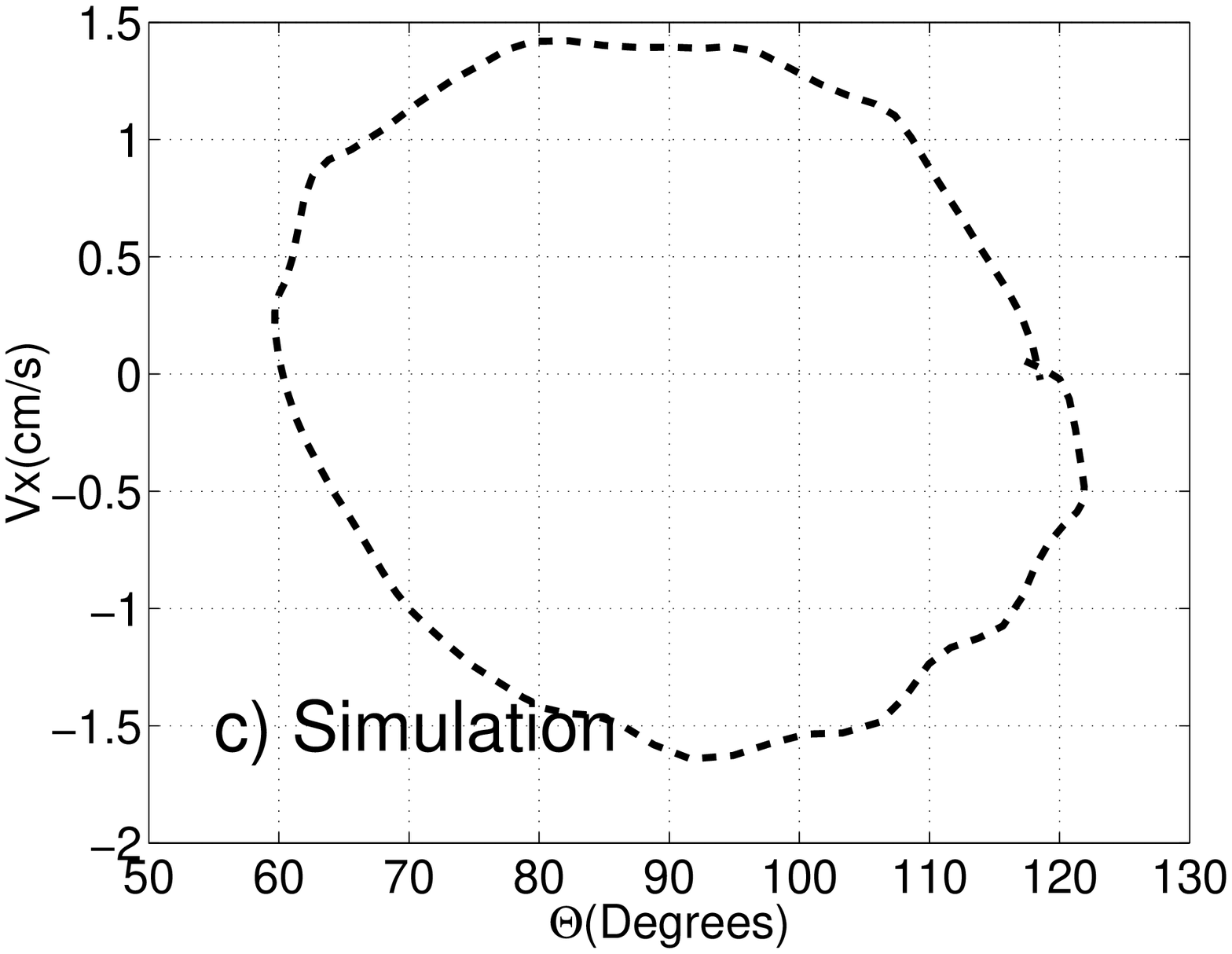,angle=0,width=7.5cm,height=5cm}
\vskip 0.5 cm
\epsfig{file=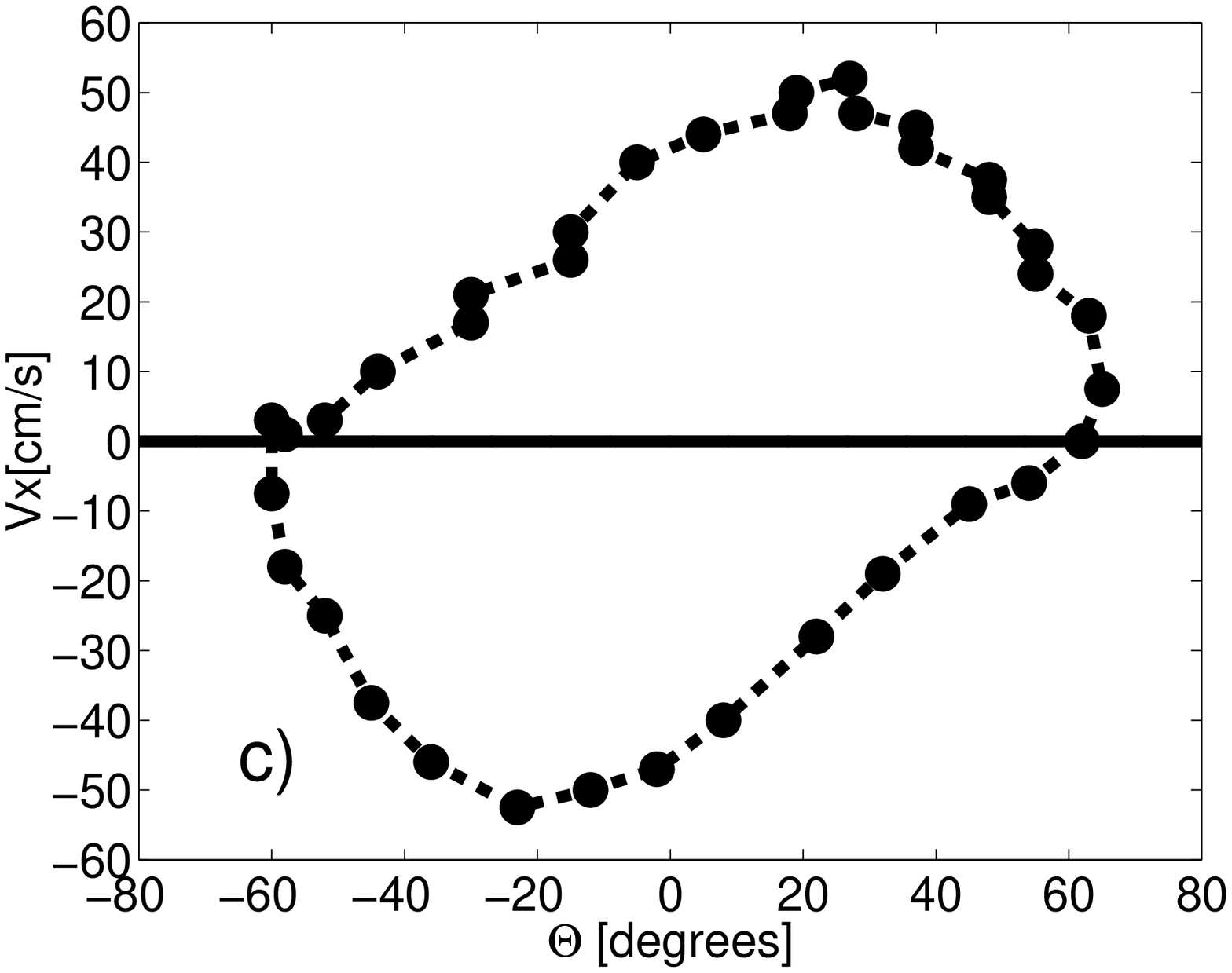,angle=0,width=7.5cm,height=5cm}

\end{center}   
 \caption{Comparison with the results of Belmonte et al. ref[12] for a falling strip 
 in the periodic regime, for the (a) Vertical orientation $\Theta$ vs time. 
 (b) Vertical velocity $V_y$ vs $\Theta$. 
 (c) Horizontal velocity $V_x$ vs $\Theta$. 
 The initial conditions are $ h_o = 96 $, $\Delta r = 0.133$, $\mu_1 = 0.025$, 
 $\theta_0 = 63.4^{o} $.}  
  
\end{figure}

In figure 12a (simulation) we show the time dependence of the vertical orientation 
with $\mu_2 = 0.025$. The value of the angular peak-to-peak amplitude is 
$\Theta_{o}=60^{o}$. The vertical velocity fig. 12b (simulation) reaches its maximal 
value $3.6\frac{cm}{seg}$ as $\Theta$ approaches $\Theta_{max}$, presenting 
a minimal drag.

The smaller vertical velocity $V_y=2.5\frac{cm}{s}$ in the oscillation is reached 
at $\Theta_{min}\sim105^{o}$. The butterfly shape of fig. 12b was also measured 
in the experimental work of Belmonte et al. \cite{Belmonte} (fig. 12b), exhibiting 
a vertical orientation $\Theta$ that oscillates at double the period of $V_y$.

The horizontal velocity oscillates around zero with the same period as $\Theta$, 
presenting its maximum value at $Vx_{max}=1.5\frac{cm}{s}$ and the minimum at 
$Vx_{min}=-1.5\frac{cm}{s}$ at $\Theta\sim90^{o}$ as seen in fig. 12c. When the 
horizontal velocity is zero the oblate takes its maximum ($120^{o}$) and minimum 
($60^{o}$) values in $\Theta$ corresponding to a non-zero value of the vertical velocity 
$V_y=3.2\frac{cm}{s}$.

\subsection{Sensitivity to a Change in the Initial Orientation In the chaotic Regime.}
\begin{figure}
 \begin{center}
\epsfig{file=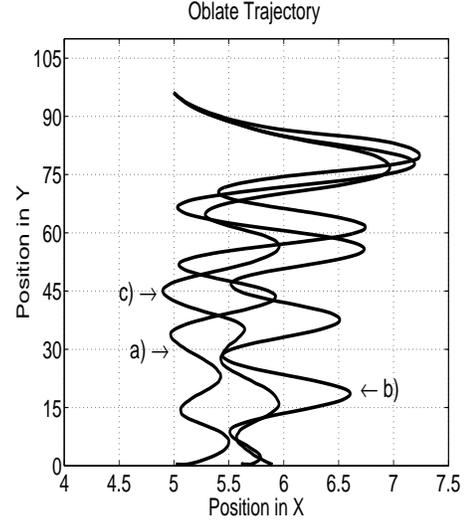,angle=0,width=6cm,height=7cm} 
\vskip 0.5 cm

\caption{Initial conditions $ h_o = 96 $, $\Delta r = 0.25$, $\mu = 0.033$ and tiny 
    variations of the initial orientation $ (a)\theta_0 = 26.6^{o} $, 
    $ (b) \theta_o= 26.6001^{o}$, $ (c) \theta_o= 26.6000001^{o}$.} 
  \end{center} 
    
\end{figure}

We can ask what is the sensitivity of the oblate to tiny changes in the initial orientation. 
Therefore we have simulated three trajectories shown in fig. 13, which have slightly different 
initial orientation. A tiny variation in the relative orientation 
($\Delta \theta_o = 10^{-3}$) produces, however, a significant variation in the shape of 
all the curves. These can be appreciated in the lower part of the trajectories. 


\begin{figure}
 \begin{center}
  \epsfig{file=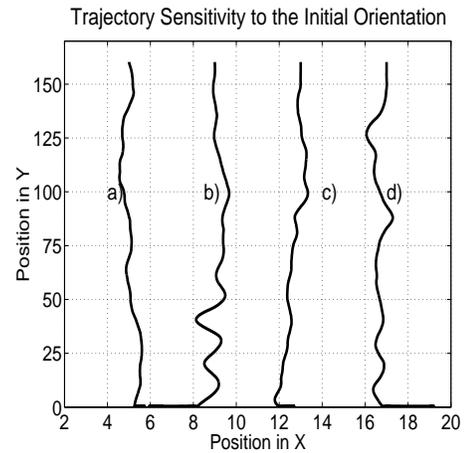,angle=0,width=6cm,height=6cm}
\vskip 0.5 cm 

   \caption{Initial conditions $ h_o = 166 $, $\Delta r = 0.25$, $\mu = 0.033$ 
    and tiny variations of the initial orientation $ (a) \theta_0 = 45.384^{o} $, 
   $(b) \theta_o= 45.033^{o}  $, $(c) \theta_o= 44.981^{o}$, $(d) \theta_o= 44.976^{o}$.} 
  \end{center} 
    
\end{figure}

In order to get better sensitivity, we have incremented the falling height to 
$h_o = 166 cm$. The resulting trajectories for four slightly different initial 
orientations in the vertical plane are presented in fig. 14. In this regime the 
system presents a high sensitivity to the initial orientation condition. For the 
four trajectories the relative angular variation is $\Delta \theta_o = 10^{-3}$. 


\begin{figure}
 \begin{center}

  \epsfig{file=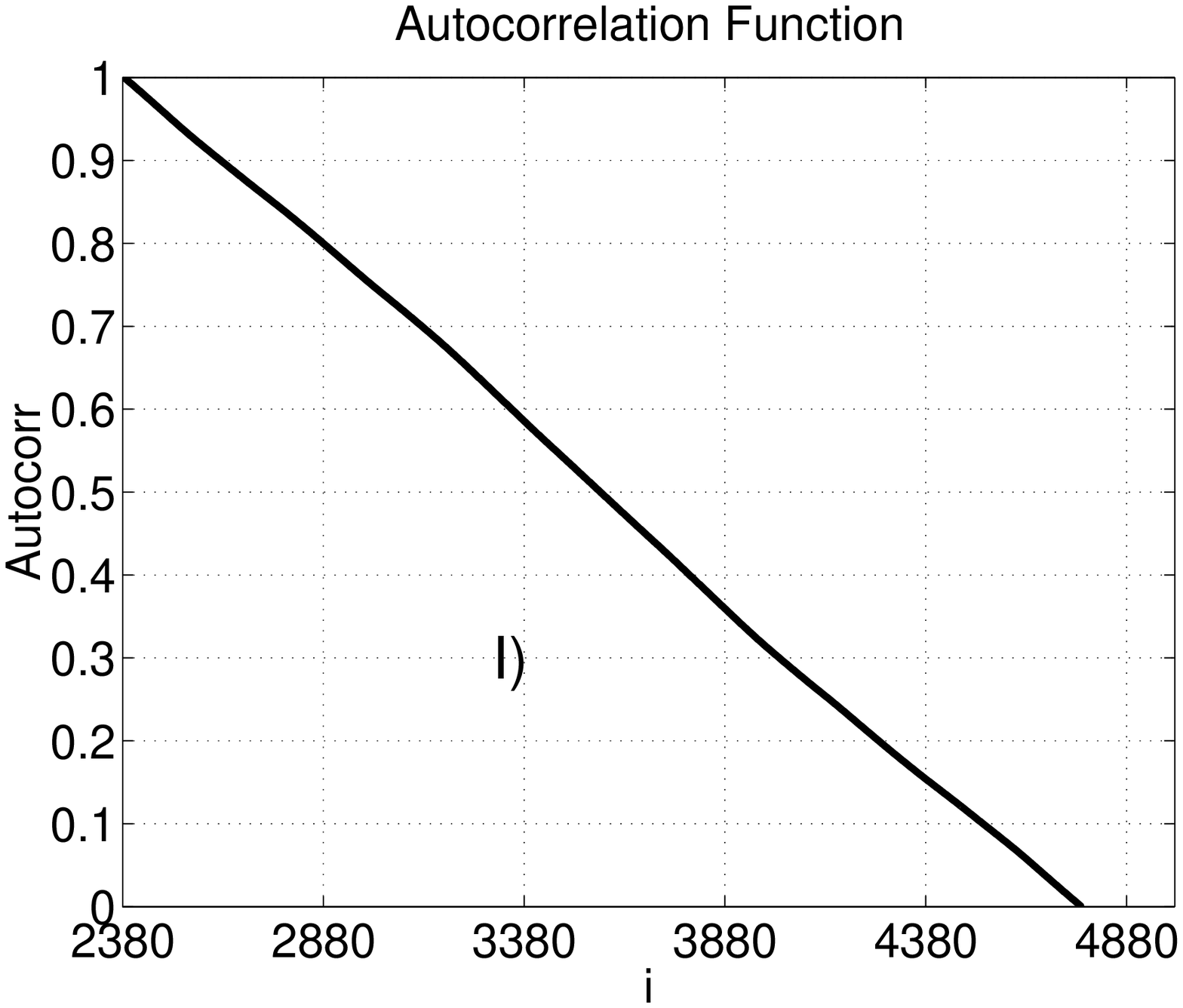,angle=0,width=7cm,height=5cm}

\vskip 0.5 cm

\epsfig{file=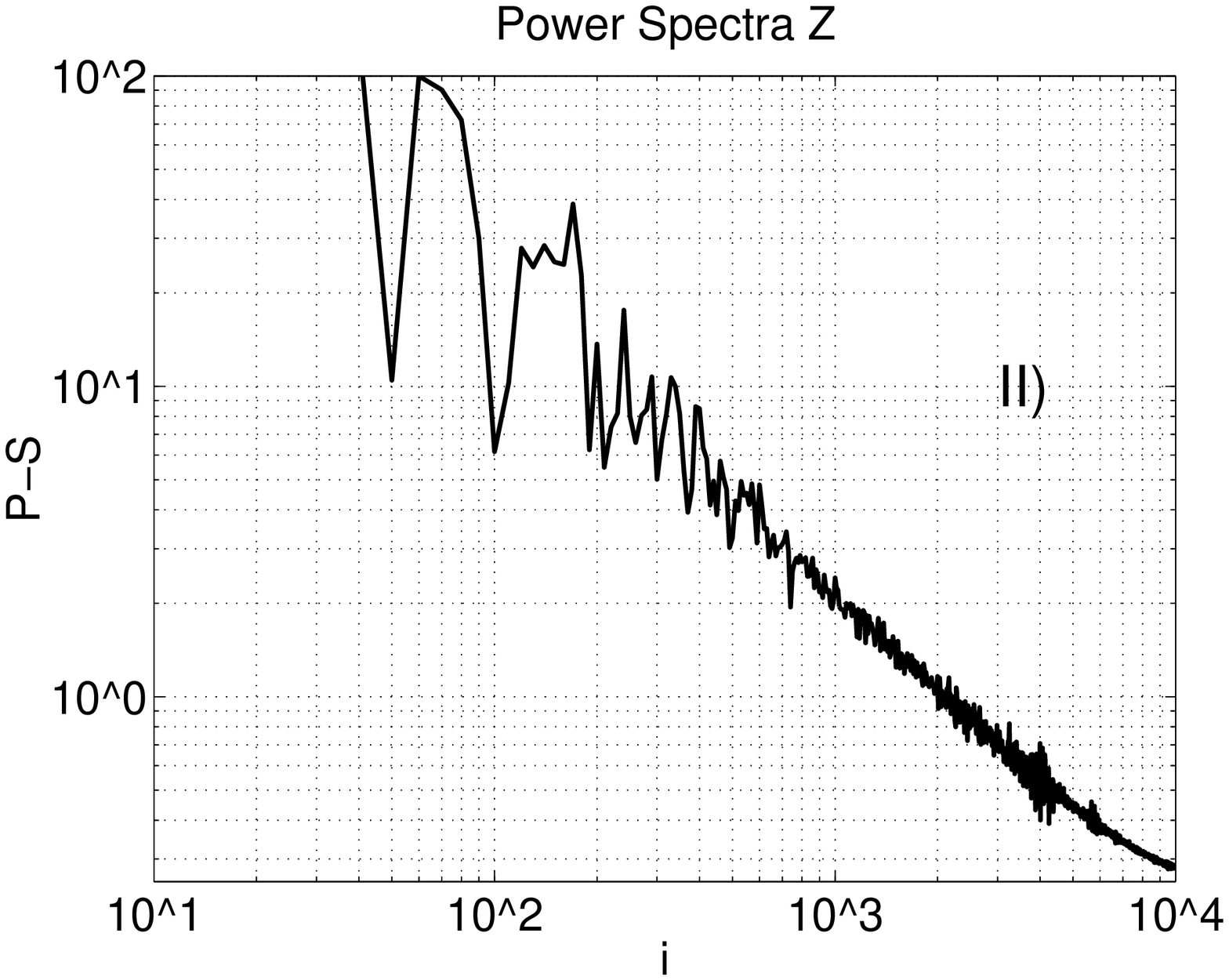,angle=0,width=7cm,height=5cm}

\vskip 0.5 cm    

  \epsfig{file=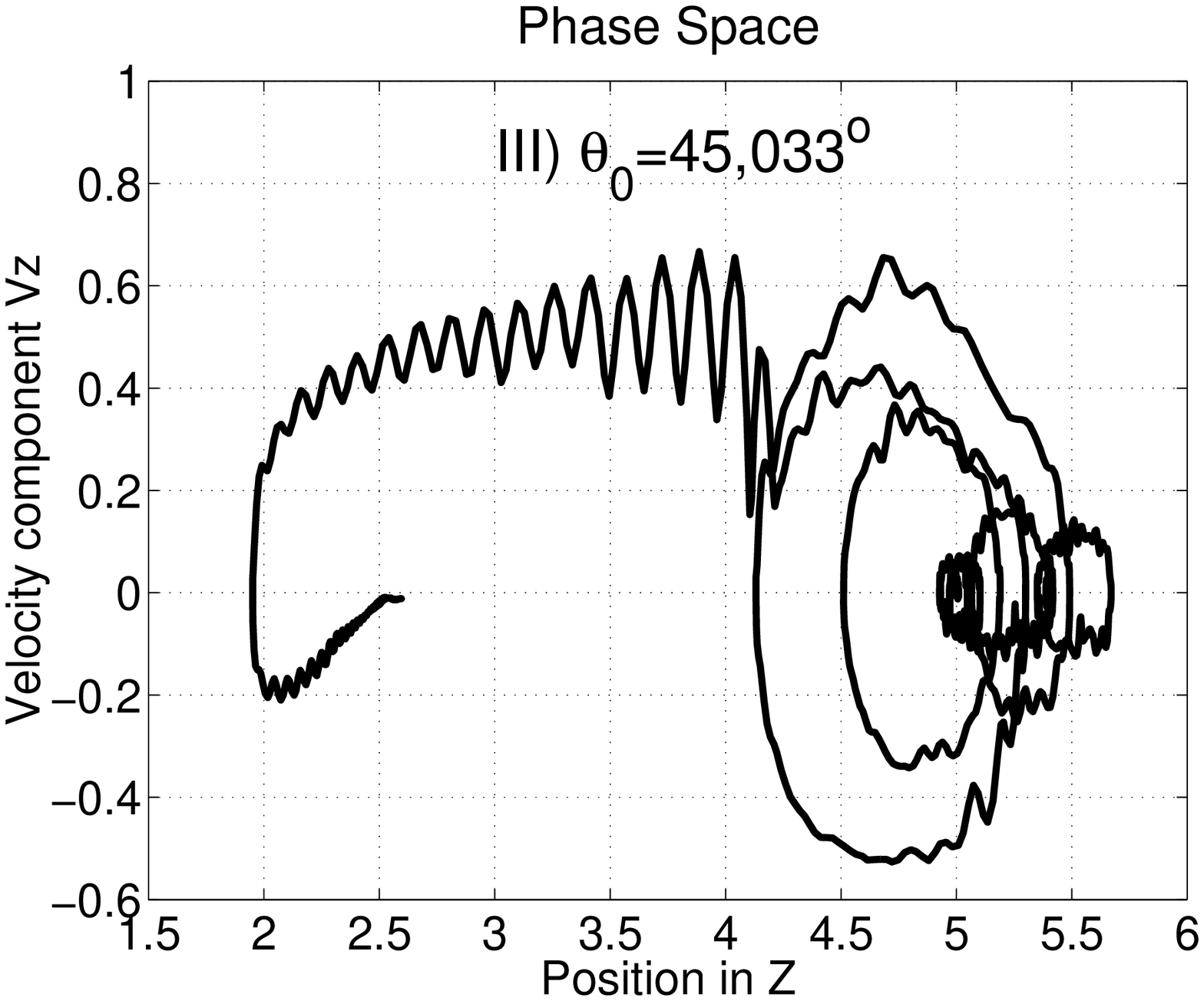,angle=0,width=7cm,height=5cm}
 
   \end{center} 
 \caption{Detection of chaos. I) Autocorrelation function for the time series 
 of $x(t)$ for the trajectory of fig. 14a. II) Power Spectra of fig. 14a III) 
 Poincare section $(p_x,x)$ for the trajectory of fig. 14a.} 
\end{figure}

Due to this dependence on small changes in the initial orientation, we proceed 
to use as a tool of diagnosis, the Fourier power spectrum of the time series of 
the horizontal coordinate $x(t), x(t+\delta t),x(t+2*\delta t) ...  $, and in our 
case $\delta t  = 0.053566$. A broad spectrum of frecuencies appears, as shown in 
fig. 15II, indicating chaotic motion.

The autocorrelation function, for the same time series (see fig. 15I), does not 
fall quickly to zero, it decreases linearly with time. The points are not 
independent of each other and a self similarity is present in the data.

In the figures 14III, we present a slices or Poincar\'{e} sections $(p_x,x)$, 
corresponding to the trajectories in fig. 14a, and which are quite irregular.

The orbits are quasi-periodic in the sense that they pass repeatedly and 
irregularly through the whole domain without ever closing on themselves, and 
without any particular time period associated with succesive passages.

The sensitivity to initial conditions is clear in these four figures. A small 
change in the initial orientation results in large changes in position and velocity. 
 

\begin{figure}
 \begin{center}
  \epsfig{file=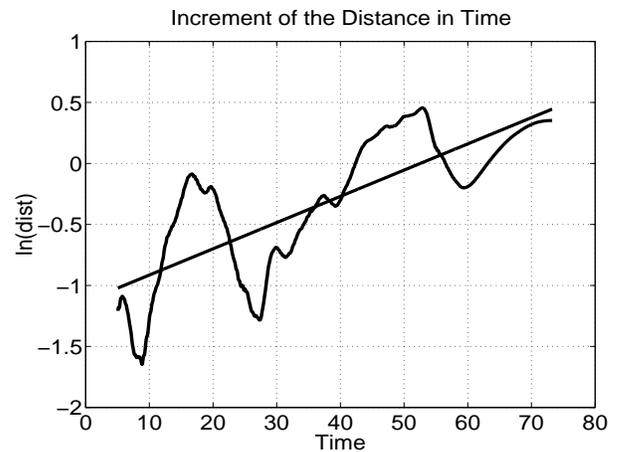,angle=0,width=8cm,height=6cm}

   \caption{Increasing logarithmic behavior for the separation distance between the 
  trajectories $(a)$ and $(c)$ in fig. 14, that slightly differ in the initial orientation angle 
  by $\Delta\Theta=0.403^{o}$.} 
 \end{center}    
\end{figure}

We can now investigate quantitatively this sensitivity by studying the increment in 
the Euclidean distance $d_{{p_1}{p_2}}=\sqrt{(x_1-x_2)^2+(y_1-y_2)^2}$, between the 
curves  presented in fig. 14 (a) and (c). Fig. 16, shows that the distance between 
nearby points has an overall exponential time dependence $ d(t) \sim \exp(\lambda t)$ 
and the fit gives an estimate for the Lyapunov exponent $\lambda = 0.052 \pm 0.005$. 
The positivity of the Lyapunov exponent is a clear indication for chaos.

\subsection{Parameter Phase Diagram.}
We explore the phase space in the dimensionless moment of inertia $I^{\star}$, which 
is the ratio of the moment of inertia of an oblate around its principal axis to the 
moment of inertia of a sphere of liquid with the same diameter and the Reynolds number 
$Re$. We do a similar analysis for our results as in the work of Field, et al. 
\cite{Nori}. It is important to remark that the mentioned experiment was for a falling 
disk, with small aspect-ratio, and we expect that the dynamics of the system will be 
close to that of an oblate ellipsoid.

The definitions of the dimensionless variables for our system are:
\begin{equation}
I^{\star}=\frac{I_{oblate}}{I_{sphere}}=\frac{5}{4}\frac{r_m}{r_M}\frac{\rho_{oblate}}{\rho_{fluid}}
=\frac{5}{4}\frac{\rho_{oblate}}{\rho_{fluid}}\Delta r
\end{equation} 

\begin{equation}
Re=\frac{U(2r_M)\rho_{fluido}}{\mu}
\end{equation}


\begin{figure}
 \begin{center}
\psfrag{This}{\LARGE $30$}

 \epsfig{file=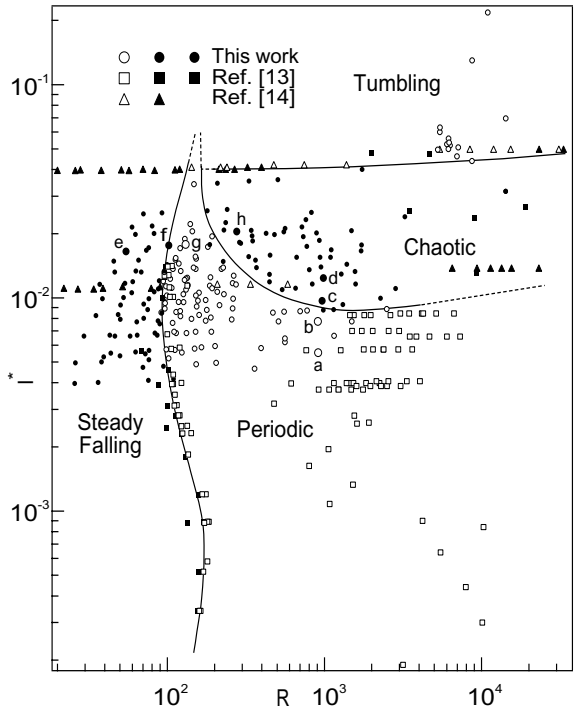,angle=0,width=8cm,height=10cm}

 \epsfig{file=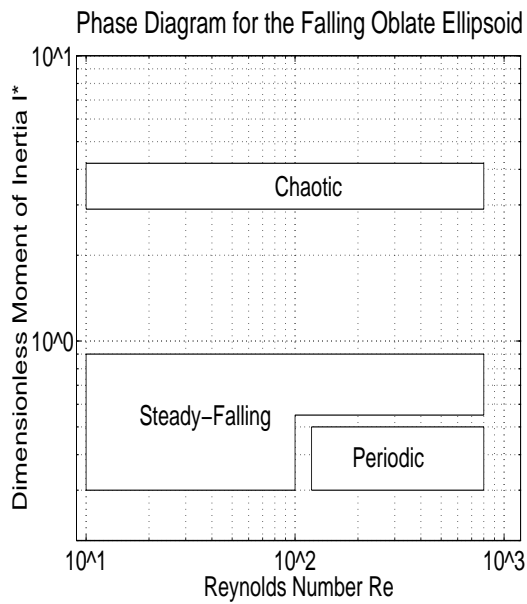,angle=0,width=7cm,height=8cm}

 \caption {The top picture shows the phase diagram of falling disks reported in ref.[11].
 In the bottom plot we present the regimes of the phase space for the falling oblate 
 obtained in our simulations.}
 \end{center} 
\end{figure}

Figure 17 bottom, shows our results. At low values of $I^{\star}$ and small Reynolds 
number (high dynamical viscosity), the left-down corner of the diagram, the motion is 
overdamped and the oblate drops to the bottom container without any oscillation. If 
the Reynolds number increases $(Re\geq100)$, fixing the moment of inertia, the trajectory 
is composed  of successive oscillations that will decrease in amplitude until the oblate 
finally comes to stop at the lowest part of the container part. For all these cases this 
type of motion was called in our results a steady-falling, and was studied in section 
IIIB-B,C,D. 

For small values of $I^{\star} \sim \Delta r \ll 1$, we have a flattened ellipsoid,
and Reynolds number $(Re\geq400)$, the trajectory, velocity and orientation are characterized 
by oscillations that repeat at equal intervals of time and space. This case was
called in our simulations the oscillatory regime, and was studied in section 
IIIB-E. 

As we increase $I^{\star}$, the object will become a sphere slightly flattened in the poles, 
and its dynamic becomes sensitive to smaller variations in the initial orientation, 
exihibiting a chaotic trajectory, which is explained in section IIIF.

If we compare our diagram with the experimental results obtained by S. Field et al, 
ref\cite{Nori} (fig. 17 top), we can see that in both pictures, the steady-falling 
region is located in the lower-left part, the oscillatory region is in the lower-right, 
and the chaotic regime is located along the top of the picture. We can say that the 
two diagrams are similar, but, with the difference, that the tumbling regime in the 
Field's diagram is not present in our results because by the comparison of the two 
diagrams in fig. 17. The tumbling regime would be found for larger dimensionsless 
moment of inertia, in our case smaller aspect-ratio, demanding a larger three dimensional 
container and impliying a very expensive computational study, than in the other 
regimes.  
 
If we take a different initial orientation angle, the new phase diagram exhibits the 
same dynamical behavior of figure 17 bottom. The coexistence of the dynamical phases, 
explained above, is independent of the initial orientation of the oblate.   

\section{Conclusions and Outlook}
The motion of a single oblate settling in a fluid in a three dimensional 
container has been studied. We found three basic regimes for the dynamics of the 
system (steady-falling, oscillatory, and chaotic). The steady-falling and the periodic 
motion exhibit a similar physical behavior as observed for flattened bodies \cite{Nori}, 
\cite{Belmonte}. With the exception that the tumbling motion is missing in our simulations. 

We have characterized the dynamics of the steady-falling regime when the dynamical 
viscosity, dropping height, and oblate's aspect-ratio are changed. Several conclusions 
can be drawn from this part of the work.

(a) The spatial trajectories $(x,y)$ are composed of oscillations that correspond to a 
damped harmonic motion. This regime is present for small values of $I^{\star}\approx0.5-1$,
$Re\approx100$ and is shown in fig.6-7. There is no variation in the trajectories when 
we increase the initial height. When the aspect-ratio is varied, the trajectories change 
very much, fig. 8. The aspect-ratio dominates strongly the type of trajectory that is present 
in the system.     

(b) The final vertical velocity $V_y$  does not depend on the initial falling height and the 
dynamical viscosity.

(c) The vertical orientation $\Theta$ of the oblate, undergoes a rotational motion till 
its mayor axis is aligned with the direction of gravity. This tendency of finding the 
minimal resistance against the fluid, is present for all Reynolds numbers in the 
range $Re\approx 30-100$ used in our simulations.

The periodic behavior in our simulations is found for ($Re\sim500$), and small 
($I^{\star}\leq0.5$). The most important characteristic on this regime is that the 
vertical orientation $\Theta$, oscillates at the double of the period of the vertical 
velocity $V_{y}$, and at the same period of the horizontal velocity $V_{x}$. This 
periodic motion is also present in the work of Belmonte et al. \cite{Belmonte}. In 
this regime the initial orientation determines the value of the oscillation amplitude in 
the spatial trajectory $(x,y)$, velocity $V_{y}$  and orientation $\Theta$. For 
$\Theta_o=90^{o}$ the amplitudes of the above quantities approach a smaller value.

The chaotic behavior is present for $I^{\star}\geq1$ and in the entire range of 
Reynolds numbers used in the simulation. The separation between the spatial 
trajectories of the falling oblate will diverge for small variations in the initial 
orientation $\Theta_{o}$, and grows exponentially in time. The value found for 
the Lyapunov exponent is around $\lambda = 0.052 \pm 0.005$. 

The construction of the phase diagram shows three well-defined dynamical regions as in the 
case of ref. \cite{Nori}. But with the difference that the chaotic behavior in the above 
reference is associated with a transition to chaos through intermittency for which we have 
no indication in our simulations. The phase diagram is independent on the initial 
orientation.   

More work to better understand the role of the fluid pressure and velocity fields 
as well as a more systematic study of the phase transition in the phase diagram seem 
necessary 

\section{Acknowledgment}

This research is part of the SFB-404-Project A7. Helpful discussions with S. Schwarzer, R. 
M\"{u}ck, and E. Kuusela are acknowledged and appreciated.

\end{multicols}
\end{document}